\documentclass[twocolumn,amsmath,amssymb,nofootinbib,floatfix]{revtex4-2}

\usepackage[pagewise]{lineno}
\usepackage{graphicx}
\usepackage{dcolumn}
\usepackage{bm}
\usepackage{hyperref}
\usepackage{color}
\usepackage{subfigure}

\newcommand{\Rmnum}[1]{\expandafter\@slowromancap\romannumeral #1@}

\def\be{\begin{equation}}
	\def\ee{\end{equation}}
\def\bea{\begin{eqnarray}}
	\def\eea{\end{eqnarray}}

\begin{document}
\title{Primordial black holes and Scalar-Induced Gravitational Waves formed by inflation potential with non-trivial characteristics}

\author{Ruifeng Zheng}
\email{
zrf2022@stu2022.jnu.edu.cn}
\author{Yanqing Xu}

\affiliation{Physics Department, College of Physics and Optoelectronic Engineering, Jinan University, Guangzhou 510632, China}

\begin{abstract}
The formation of primordial black holes (PBHs) generally requires large density perturbations, which is widely supported by researchers. This paper studies the local coupling properties of the Starobinsky and Kachru-Kallosh-Linde-Trivedi (KKLT) potentials by introducing a linear Lorentzian-type coupling, which locally breaks the slow-roll conditions. We find that both positive and negative couplings can generate a considerable abundance of PBH. Additionally, we study the scalar-induced gravitational waves (SIGWs) generated by this model.
\end{abstract}

\maketitle	

\section{INTRODUCTION}\label{(1)}
\par
Black holes, as one of the most extreme predictions of Einstein's theory of general relativity, have long inspired humanity to explore the deepest mysteries of the universe. Conventionally, black holes are understood to form primarily through the gravitational collapse of massive stellar cores (typically exceeding 20 solar masses) during supernova explosions at the end of stellar evolution, or via the merger and accretion of stellar-mass or intermediate-mass black holes. However, within the extreme high-energy conditions prevalent in the very early universe, a distinct class of black holes with an alternative origin may exist: Primordial Black Holes (PBHs) \cite{Carr:1974nx, Carr:2026hot, Khlopov:2008qy}. These types of black holes do not originate from the death of stars, but are directly born from the strong non-uniformity of energy distribution (density perturbations) in the very early universe (typically within the first second after the Big Bang). Their formation mechanism is thus decoupled from standard stellar evolution. Theoretically, PBHs can possess a mass spectrum spanning an extraordinarily wide range, from subatomic scales ($\sim10^{-5}g$) to supermassive scales ($>10^{5}M_\odot$) and beyond.

PBHs were first systematically discussed by Hawking \cite{Hawking:1971ei, Carr:1974nx} and Zeldovich \cite{Zeldovich:1967lct} in the 1960s and 1970s. The core formation mechanism posits that during the radiation-dominated epoch immediately following inflation, if the density perturbation amplitude of certain spatial regions is large enough to exceed the critical value ($\delta>\delta_c$), such that their self-gravity overcomes the cosmic expansion and radiation pressure, then these regions will undergo direct gravitational collapse, skipping the long process of star formation and directly forming black holes. This gravitational collapse is thought to occur at cosmic times $t<1s$. Consequently, PBH formation depends critically on the spectral characteristics of primordial density perturbations in the early Universe, particularly at small scales. The origin of these perturbations is usually closely related to the details of inflation models, cosmic phase transitions (such as QCD phase transitions), the generation of cosmic strings or other topological defects, and the evolution of scalar fields. Therefore, various PBH formation mechanisms have been proposed \cite{Garcia-Bellido:2017mdw, Ballesteros:2017fsr, Carr:2017edp, Pi:2017gih, Cai:2018tuh, Ballesteros:2018wlw, Pi:2019ihn, Fu:2019ttf, Chen:2019zza, ZhengRuiFeng:2021zoz, Carr:2018poi, Drees:2019xpp, Anguelova:2020nzl, Lin:2020goi, Qiu:2022klm, Choudhury:2024aji, Choudhury:2024jlz, Mahbub:2019uhl, Ashoorioon:2020hln, Karam:2022nym, Saini:2017tsz, Wang:2025hwc, Maiti:2025ijr, Kouniatalis:2025orn}.

PBHs within certain mass ranges are regarded as plausible candidates for dark matter \cite{Belotsky:2014kca, Belotsky:2018wph}. If PBHs constitute all or a significant part of dark matter, this could offer new perspectives on the fundamental properties of dark matter, avoiding the dilemma of traditional weakly interacting massive particles (WIMPs) waiting to be selected as particles that have not yet been discovered in direct detection and accelerator experiments. However, as one of the candidates for dark matter, PBHs are not without challenges. Various astronomical observations and cosmological limitations, such as the cosmic
microwave background (CMB) and big bang nucleosynthesis (BBN) \cite{Serpico:2020ehh, Carr:2020gox, Acharya:2020jbv}, BH evaporation \cite{Carr:2009jm}, Gravitational-Wave Lensing (GW-Lensing) \cite{Jung:2017flg}, gamma-ray emission \cite{Laha:2019ssq, Dasgupta:2019cae, Cai:2020fnq, Tan:2022lbm}, gravitational waves \cite{Wong:2020yig, Kavanagh:2018ggo, Kimura:2021sqz}, etc. impose stringent constraints on the abundance of PBHs as dark matter candidates across different mass scales. These constraints indicate that if PBHs exist, they are unlikely to dominate dark matter in all mass ranges. However, they may plausibly constitute either the entirety or a substantial fraction of dark matter within specific, currently observationally viable mass windows, or exist as a subcomponent of dark matter.

Ultra-slow-roll (USR) inflation is considered one of the effective mechanisms for the formation of PBHs \cite{Garcia-Bellido:2017mdw, Nandi:2025ihs}. A viable USR model must simultaneously satisfy cosmological constraints on large scales while significantly amplifying the primordial power spectrum at small scales, which can be achieved by a local inflection point in the potential \cite{Mishra:2019pzq}. In addition, large scalar perturbations may contribute to the generation of Scalar-Induced Gravitational Waves (SIGWs) \cite{Baumann:2007zm, Ananda:2006af, Domenech:2021ztg, Iovino:2025xkq, Kohri:2018awv}, which imposes further constraints on the enhancement of the primordial power spectrum during inflation.

In this paper, we study a toy inflation model characterized by the local coupling of the base inflationary potential with the Lorentzian function. This special mechanism of locally coupled dynamic adjustment allows the scalar field to transition from conventional slow-roll (SR) inflation to USR inflation over a certain period. The SR parameters exhibit exponential changes, which lead to a sharp increase in the power spectrum and can potentially satisfy the requirements for PBH formation. Our results indicate that both positive and negative coupling lead to a significant enhancement of the power spectrum on small scales. In addition, we discuss the SIGWs produced by the model, which could be detected by current or future gravitational wave experiments.

This paper is organized as follows: In Sec. \ref{S2}, we briefly review the relevant details of SR and USR inflation. In Sec. \ref{S3}, we introduce the Starobinsky and KKLT potentials with a local coupling; the numerical results indicate that the power spectrum satisfies CMB constraints on large scales while exhibiting enhanced peaks at small scales. In Sec. \ref{S4}, we calculate the abundance of PBHs formed in our model. In Sec. \ref{S5}, we discuss the SIGWs generated by scalar perturbations, and In Sec. \ref{S6}, we present our conclusions and discussions. In the appendix, we also provide a simple extension of this model, whose predicted signal could be detected by gravitational wave experiments in different frequency bands.

\section{SR and USR of scalar field}\label{S2}
\par
The action of a single scalar field inflation model is usually given by 
\begin{equation}
\begin{aligned} \label{action}
S=\int d^{4} x \sqrt{-g}\left[\frac{M_{p l}^{2} R}{2}-\frac{1}{2} \partial_{\mu} \phi \partial^{\mu} \phi-V(\phi)\right]~,
\end{aligned}
\end{equation}
where $R$ is the Ricci scalar and $V(\phi)$ is the inflationary potential. The corresponding Friedman equation and the equation of motion for the scalar field can be written as
\begin{equation}
\begin{aligned} \label{potential}
H^{2}=\frac{1}{3 M_{p l}^{2}}\left(\frac{1}{2} \dot{\phi}^{2}+V(\phi)\right)~,\\
\ddot{\phi}+3 H \dot{\phi}+V^{\prime}(\phi)=0~,
\end{aligned}
\end{equation}
where $H$ is the Hubble parameter and $ \dot{\phi}=d\phi/dt$. To describe the inflationary dynamics, the SR parameter is usually defined as
\begin{equation}
\begin{aligned}\label{SR}
\epsilon_H &= -\frac{\dot{H}}{H^2}~,
\\
\eta_H &= -\frac{\ddot{H}}{2\dot{H}H}~,
\\
\xi_H &= \frac{\ddot{H}}{2H^2\dot{H}} - 2\eta_H^2~.
\end{aligned}
\end{equation}
SR inflation requires the SR parameters to satisfy $\epsilon_H, \eta_H \ll 1$ throughout inflation. In addition, to solve the horizon and flatness problems of the standard Big Bang cosmology,  the total number of e-foldings is required \cite{Baumann:2009ds}
\begin{equation}
\begin{aligned}\label{EF}
N \equiv N_{i} - N_{e} = \int_{t_{i}}^{t_{e}} H \left( t \right) dt \ge  60~.
\end{aligned}
\end{equation}
\par
Now let us discuss the perturbations generated during inflation. In the context of the FRW metric, the Mukhanov–Sasaki (MS) variable is defined as $u\equiv z\zeta~$, with $z\equiv a\dot{\phi } /H $, and it satisfies the Mukhanov-Sasaki equation \cite{Sasaki:1986hm, Mukhanov:1988jd}
\begin{equation}
\begin{aligned}\label{MS}
u_{k}^{\prime \prime} + \left( k^{2} - \frac{z^{\prime \prime}}{z} \right) u_{k} = 0~,
\end{aligned}
\end{equation}
where the prime denotes the derivative with respect to comoving time $\tau$, and the effective potential term is given by
\begin{equation}
\begin{aligned}\label{EP}
\frac{z^{\prime \prime}}{z}=2 a^{2} H^{2}\left(1+\epsilon_{H}-\frac{3}{2} \eta_{H}+\epsilon_{H}^{2} \right. \\
\left.
+\frac{1}{2} \eta_{H}^{2}-2 \epsilon_{H} \eta_{H}+\frac{1}{2} \xi_{H}\right)~,
\end{aligned}
\end{equation}
where $a$ is the scale factor. By solving Eq. (\ref{MS}) and using the relation between the MS variable and the dimensionless primordial power spectrum \cite{Ballesteros:2017fsr}, we obtain
\begin{equation}
\begin{aligned}\label{ps}
P_{S}=\left.\frac{k^{3}}{2 \pi^{2}} \frac{\left|u_{k}\right|^{2}}{z^{2}}\right|_{k \ll a H}~.
\end{aligned}
\end{equation}
In addition, under the SR approximation, the perturbation power spectrum can be written as 
\begin{equation}
\begin{aligned}\label{PS}
P_{S}=\frac{1}{8 \pi^{2} \epsilon_{H}}\left(\frac{H}{M_{p l}}\right)^{2}~.
\end{aligned}
\end{equation}
Large-scale observations indicate that the primordial power spectrum has near scale invariant, which means the scalar spectral tilt \cite{Planck:2018jri}
\begin{equation}
\begin{aligned}\label{PZS}
n_s=1+2\eta_H-4\epsilon_H\sim0.968~,
\end{aligned}
\end{equation}
and the tensor-to-scalar ratio
\begin{equation}
\begin{aligned}\label{TSR}
r\approx16\epsilon_H<0.03~.
\end{aligned}
\end{equation}
Planck 2018 constrains the amplitude of the primordial power spectrum to $\sim10^{-9}$ on CMB scales \cite{Planck:2018jri}, which is insufficient for PBH formation. To generate a sufficient abundance of PBHs, the power spectrum must be enhanced to $\sim10^{-2}$ on small scale \cite{Motohashi:2017kbs}. A reasonable way is to locally break the SR condition, thereby entering an USR state. In the USR regime, the potential becomes sufficiently flat that $V'(\phi) \approx 0$, and the equation of motion for the scalar field reduces to \cite{Martin:2012pe, Kinney:2005vj}
\begin{equation}
\begin{aligned}\label{USR}
\ddot{\phi } +3H\phi = -V'(\phi) = 0~,
\end{aligned}
\end{equation}
which implies that the SR parameter $\epsilon_H$ decays exponentially. According to Eq. (\ref{PS}), this leads to an exponential growth of the power spectrum. It is important to note that this evolution violates the SR condition, causing the SR parameter $\eta_H=3+\epsilon_H>1$ \cite{Di:2017ndc}. As a result, the SR approximation breaks down and the system enters a USR phase. However, this phase is only temporary. Once the scalar field exits the local coupling region, the system resumes SR inflation and continues until inflation ends.
\section{Local Coupling in the Inflationary Potential}\label{S3}
The SR inflationary scenario can successfully account for current observational constraints, such as the scalar spectral index $n_s\approx0.968$ and the tensor-to-scalar ratio $r<0.03$. To achieve a USR phase, one approach is to construct a new USR potential, this requires a comprehensive construction of potential \cite{Garcia-Bellido:2017mdw}. A simpler alternative is to introduce a local feature into a known SR potential \cite{Mishra:2019pzq}. This latter method temporarily violates the SR conditions on small scales while remaining consistent with large-scale observations.

Here we adopt a method for realizing localized modifications by coupling a Lorentzian function to the potential. This coupling induces inflection points at suitable locations, leading to a pronounced peak in the power spectrum on small scales. It is worth emphasizing that this method of generating inflection points by coupling Lorentzian functions is universal: in principle, it can be locally coupled to any inflationary potential without significantly affecting large-scale observational constraints. In this paper, we first adopt the Starobinsky potential as the basic potential \cite{Starobinsky:1980te}. Its full form is given by
\begin{equation}
\begin{aligned}\label{Potential}
V(\phi)=V_0\left ( 1-e^{A_0\phi} \right )  ^2\left ( 1\pm \frac{B_1C_1}{(\phi-D_1)^2+C_1^2}  \right )~,
\end{aligned}
\end{equation}
where $V_0$ fixes the overall CMB normalization, and $A_0,B_1,C_1,D_1$ represent the relevant parameters. Specifically, $A_0$ determines the specific form of the basic potential, while $B_1, C_1, D_1$ represent the position, width, and amplitude of inflection point, respectively. The ‘‘$+$’’ sign denotes positive coupling, and ‘‘$-$’’ sign denotes negative coupling. To satisfy large-scale observational constraints, we choose $V_0=10^{-10}M_{pl}^4$ and $A_0=-0.8165M_{pl}^{-1}$, the initial value of the scalar field is set to $\phi_i=5.4M_{pl}$, which conforms to the constraints of current cosmological observations
\begin{equation}
\begin{aligned}
n_s&=0.966338~,\\
r&=0.00323652~.
\end{aligned}
\end{equation}
Although the coupling term can also affect the two parameters mentioned above ($n_s$ and $r$), it should be noted that the coupling term only affects the local potential. and its impact on CMB scales is negligible.

\begin{figure}[!htbp]
	\centering
	\includegraphics[scale=0.25]{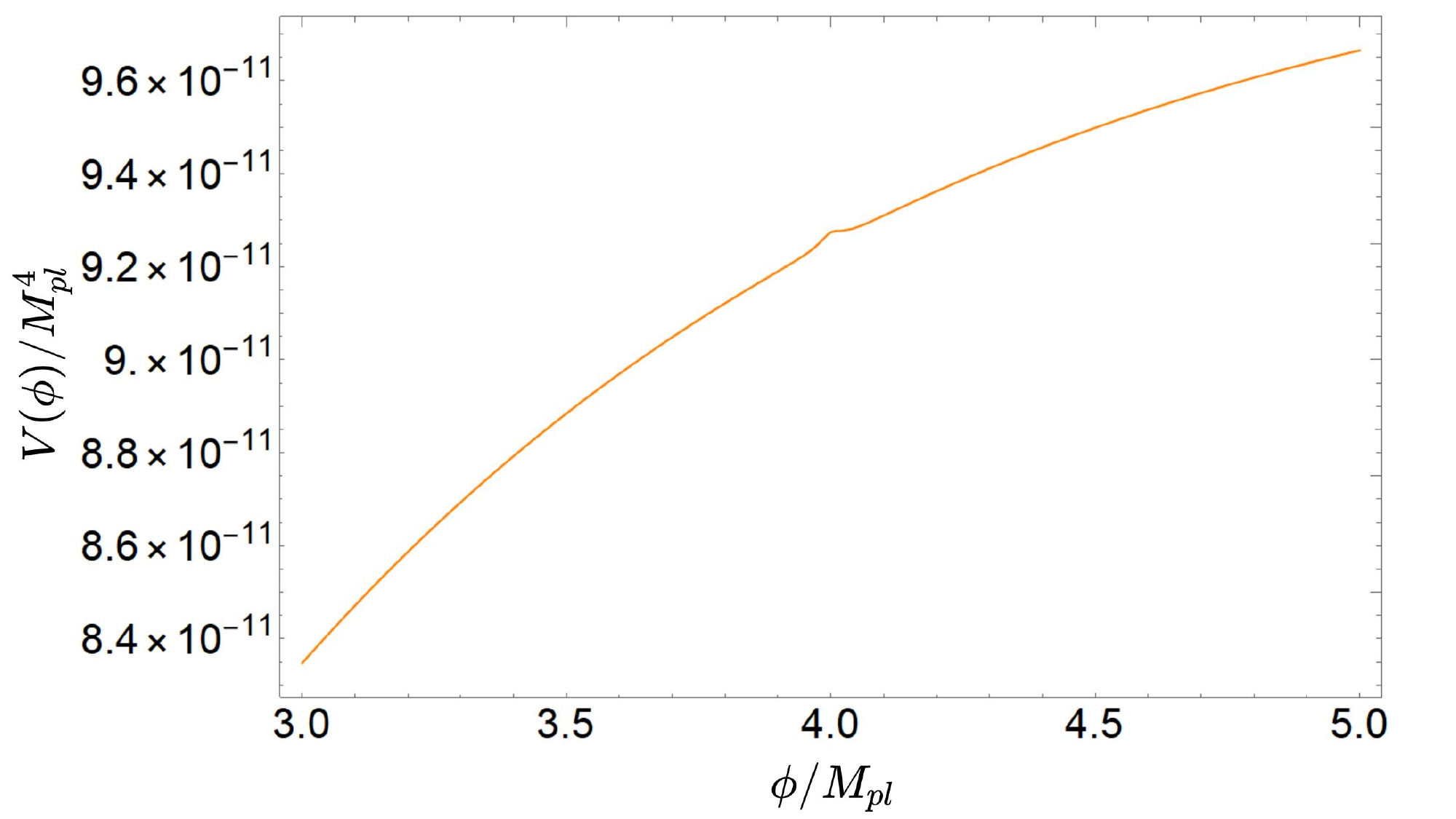}
	\caption{The Starobinsky potential with local positive coupling is given by Eq. (\ref{Potential}), with the parameters set as in Eq. (\ref{CS1}). The coupling is localized at $\phi_1=4M_{pl}$, manifesting as a local bump. The initial scalar field value is set to $\phi_i=5.4M_{pl}$.}
	\label{SNTQ}
\end{figure}

Firstly, we consider positive coupling ‘‘$+$’’ with the following parameter values:
\begin{equation}
\begin{aligned}\label{CS1}
B_1&=6.23278\cdot10^{-5}M_{pl}~,\\
C_1&=0.025M_{pl}~,\\
D_1&=4M_{pl}~.
\end{aligned}
\end{equation}
The corresponding potential is shown in Fig.~\ref{SNTQ}, we can see that the effect of positive coupling is reflected in $\phi_1=4M_{pl}$, manifesting as a local bumpy of potential, this creates a locally flat range of potential. The width and height of the bump directly affect the peak of the power spectrum, and ultimately affect the abundance of the PBHs formed. Therefore, fine-tuning of these parameters is required. 
\par
\begin{figure}[!htbp]
	\centering
	\includegraphics[scale=0.25]{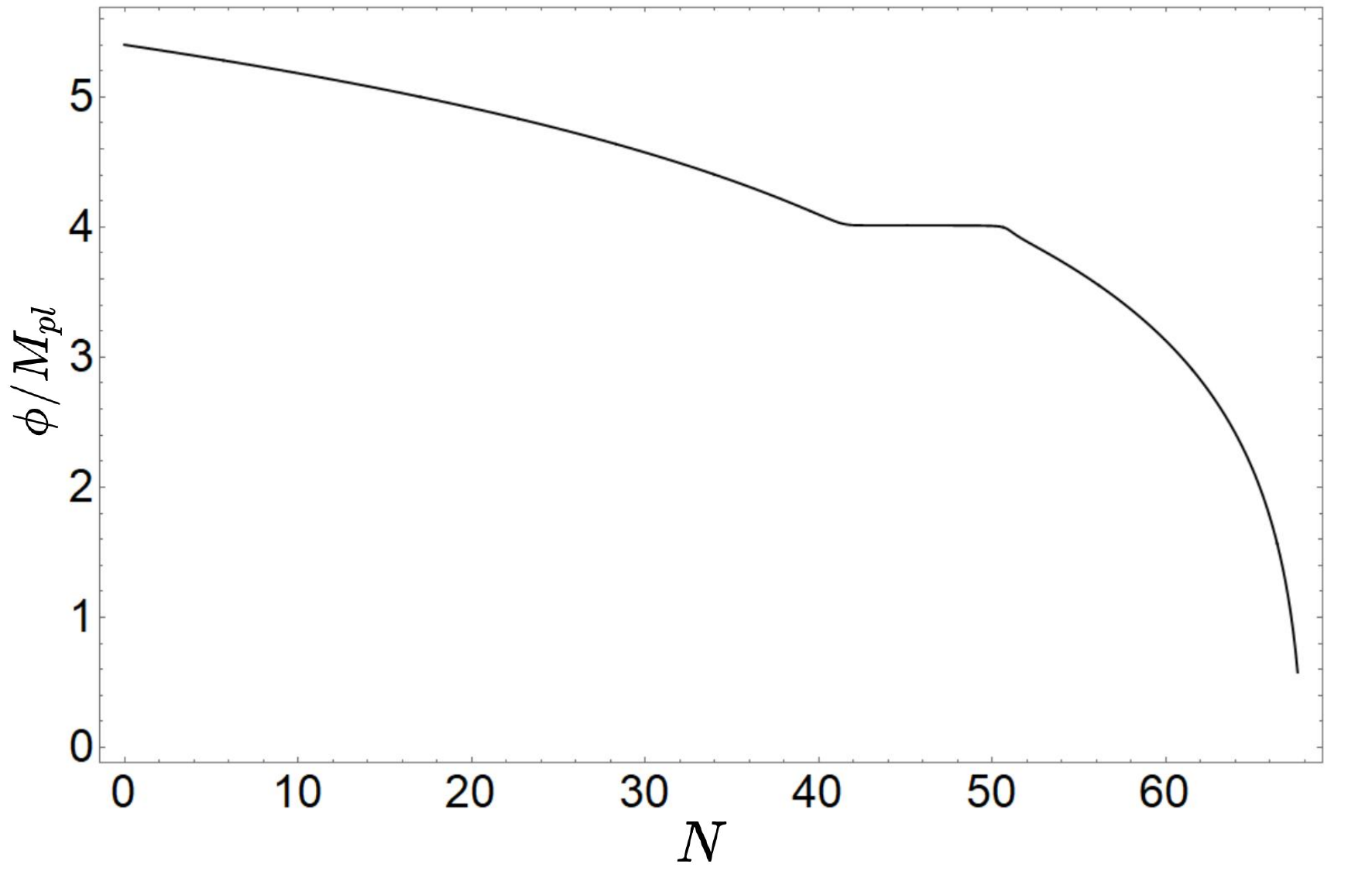}
	\caption{The figure shows the evolution of the scalar field $\phi$ as a function of the e-foldings number $N$ for the Starobinsky potential with a local positive coupling. At $N\approx45$, the scalar field transitions from SR inflation to USR inflation.}
	\label{CFTQ}
\end{figure}

As shown in Fig.~\ref{CFTQ}, by solving the equation of motion (\ref{potential}) for the scalar field $\phi$, we can obtain the evolution curve of the scalar field with respect to the e-foldings number $N$. Due to the presence of local coupling, the SR condition is temporarily violated near the bump. the motion of the scalar field at this bump is extremely slow, causing the system to transition from SR to USR dynamics. 
\par
\begin{figure}[!htbp]
	\centering
	\includegraphics[scale=0.28]{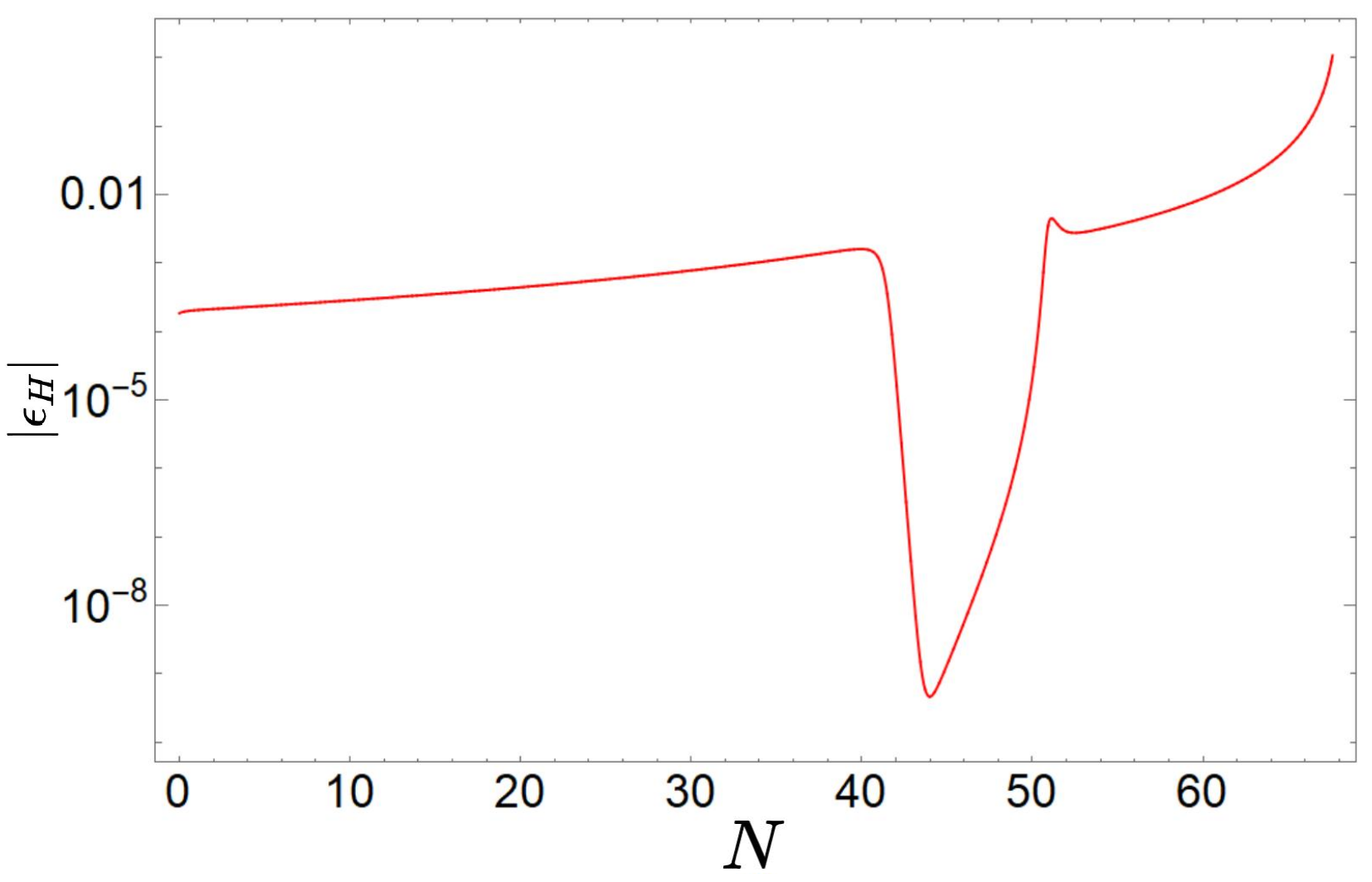}
	\caption{The figure shows the evolution of the SR parameter $\epsilon_H$ as a function of the e-foldings number $N$ for the Starobinsky potential with a local positive coupling. At $N\approx45$, $\epsilon_H$ shows a sharp downward trend.}
	\label{MGTQ}
\end{figure}
\begin{figure}[!htbp]
	\centering
\includegraphics[scale=0.28]{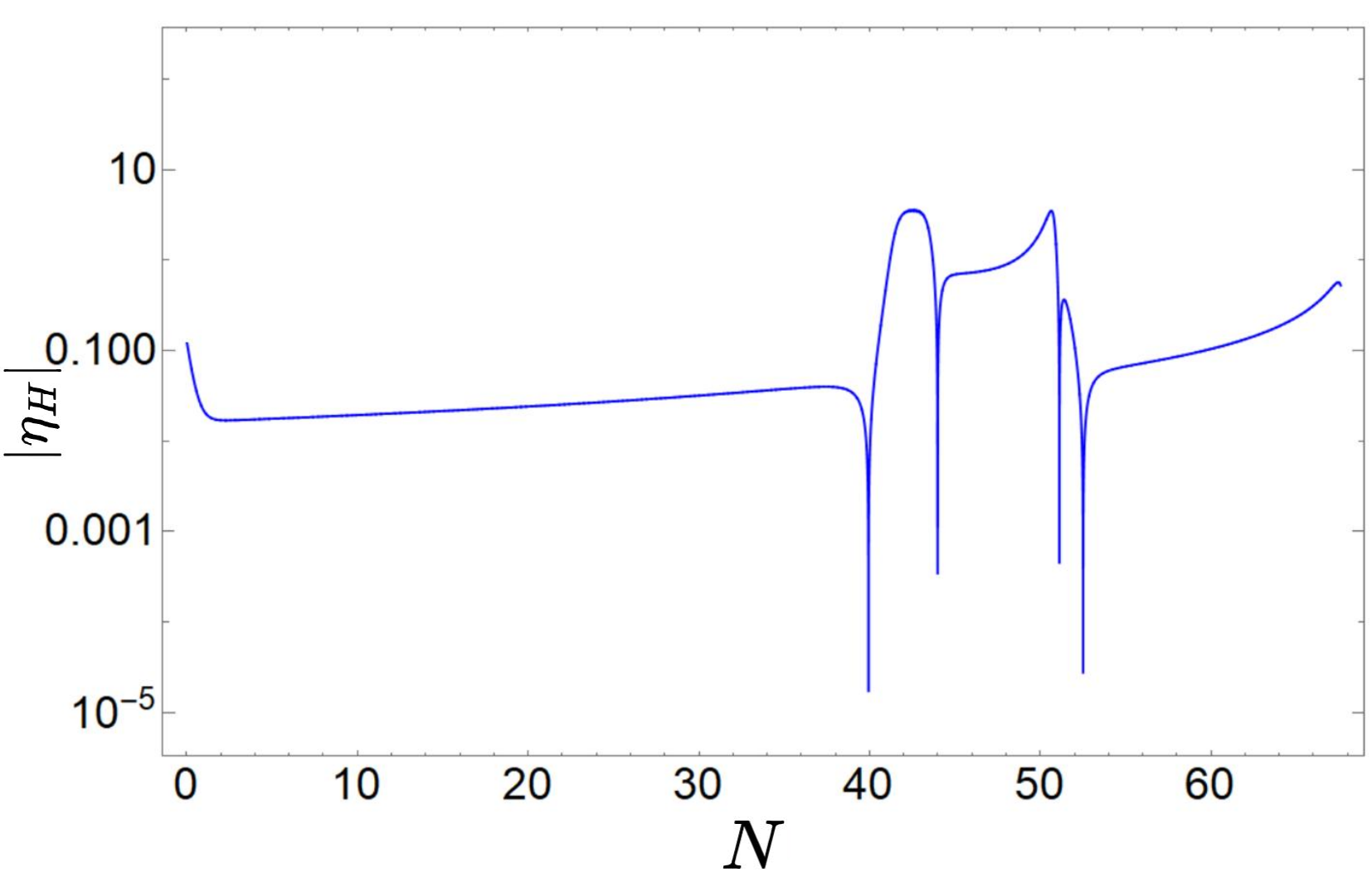}
	\caption{The figure shows the evolution of the SR parameter $\eta_H$ as a function of the e-foldings number $N$ for the Starobinsky potential with a local positive coupling. At $N\approx45$, $\eta_H$ temporarily violates the SR condition.}
	\label{MGTQ2}
\end{figure}

The coupling induces significant changes in the local evolution of the SR parameters $\epsilon_H$ and $\eta_H$. As shown in Fig.~\ref{MGTQ} and Fig.~\ref{MGTQ2}, $\epsilon_H$ drops sharply near $N\approx45$, while $\eta_H$ increases and violates the SR condition $\eta_H>1$, which is precisely one of the characteristics of USR inflation. However, this doesn't mean the end of inflation. After the scalar field passes through the localized coupling, it re-enters the SR regime, again satisfying $\epsilon_H, \eta_H < 1$, and eventually end inflation at $N \approx 67$.
\par
\begin{figure}[!htbp]
	\centering
\includegraphics[scale=0.19]{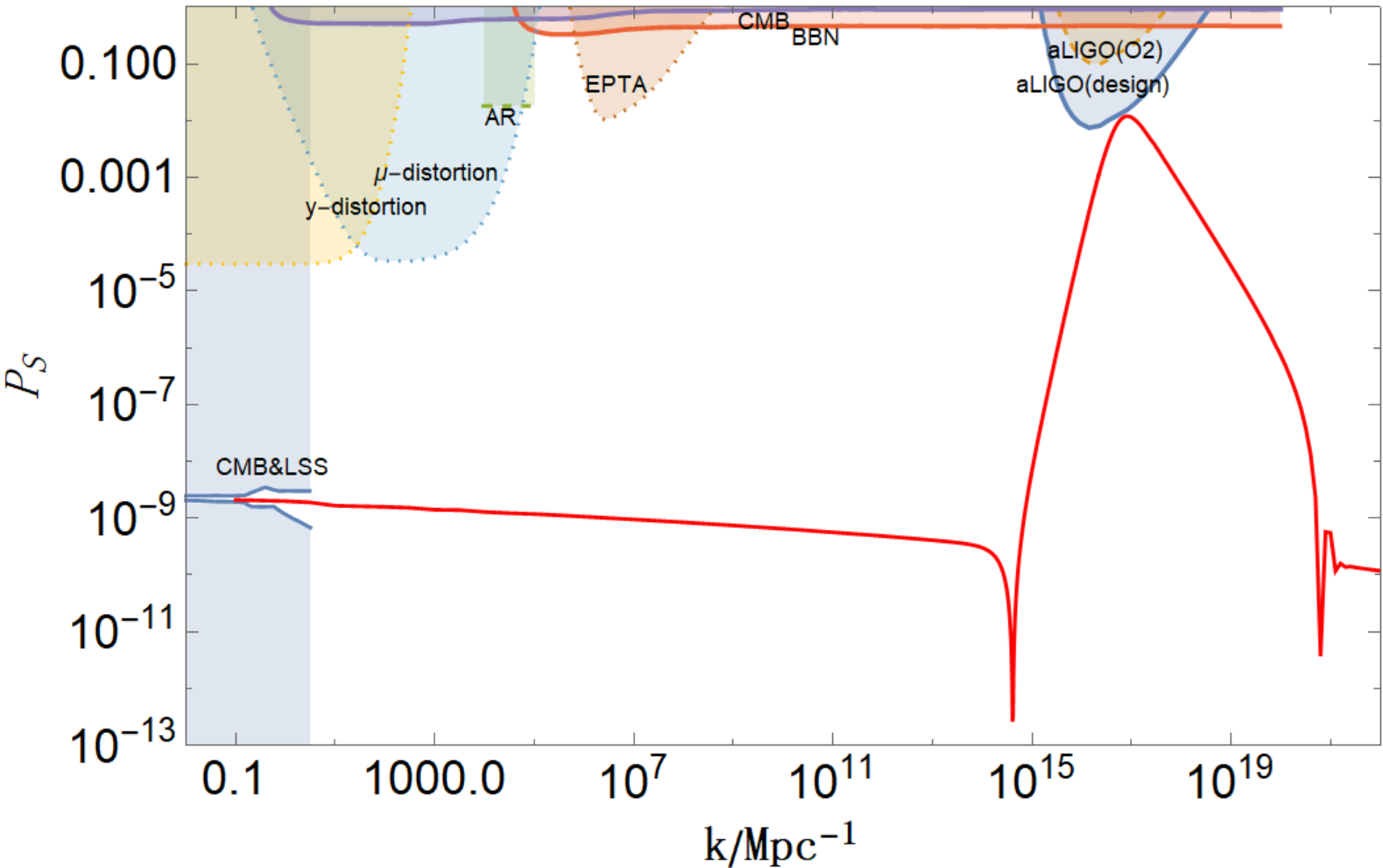}
	\caption{The relationship between power spectrum with positive coupling of Starobinsky potential and the wave-number $k$. The color region is excluded by the current observation, our numerical results are consistent with the observational constraints.}
	\label{GLPTQ}
\end{figure}

By numerically solving the MS equation (\ref{MS}), we obtain the relationship between powers pectrum $P_S$ and the wave-number $k$. As shown in Fig.~\ref{GLPTQ}, on large scales the power spectrum is consistent with CMB observational constraints ($\sim10^{-9}$), while at small scales $k\sim10^{17}\mathrm{Mpc^{-1}}$, it exhibits a pronounced peak ($\sim10^{-2}$) in the power spectrum.
\par
\begin{figure}[!htbp]
	\centering
	\includegraphics[scale=0.27]{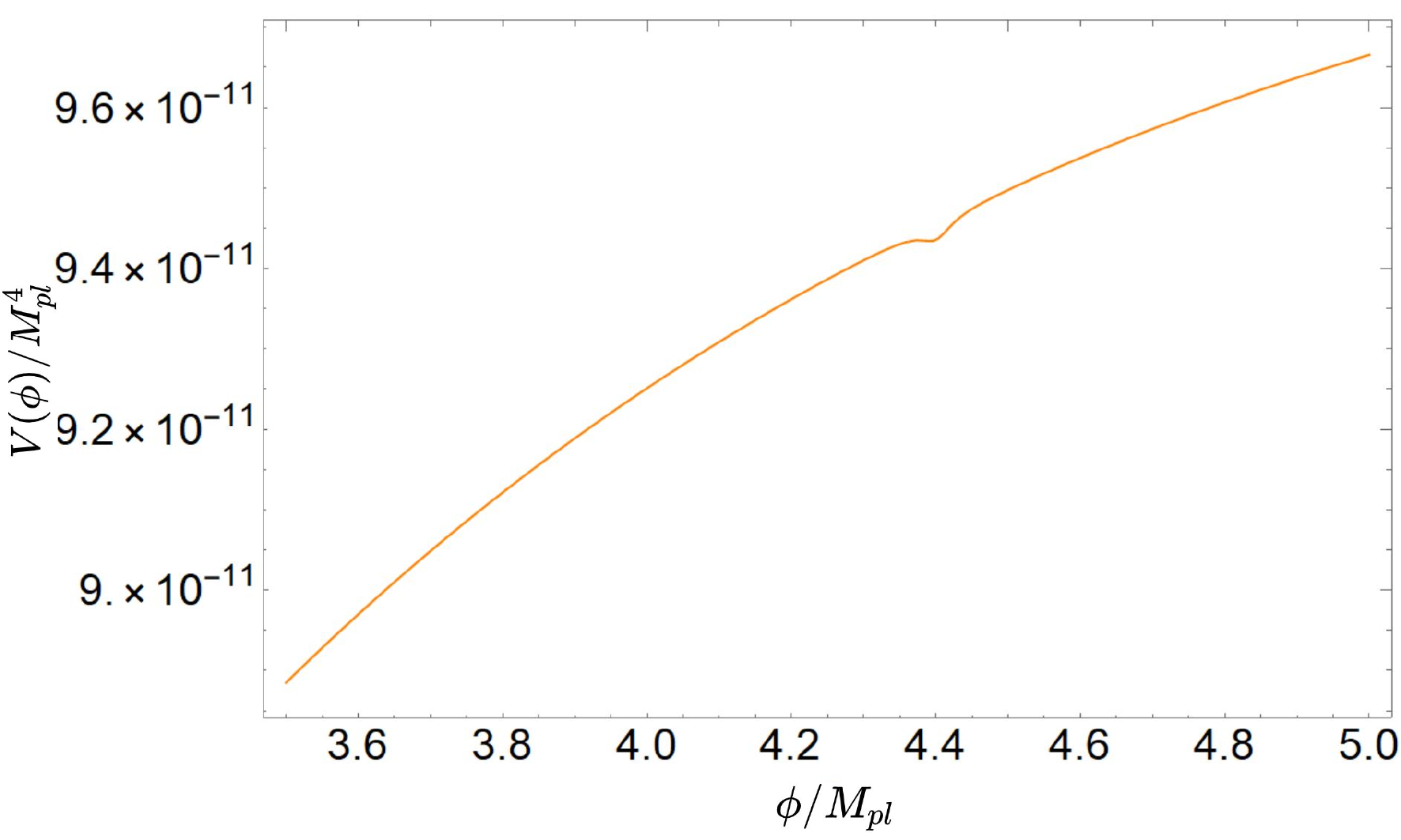}
	\caption{The Starobinsky potential with local negative coupling, and the parameter values correspond to Eq. (\ref{NC}), where the coupling is localized at $\phi_2=4.4M_{pl}$, manifesting as a local dip. The initial value of the scalar field is at $\phi_i=5.4M_{pl}$.}
	\label{SNOX}
\end{figure}

In addition to positive coupling for enhancing the small-scale power spectrum, we demonstrate that negative coupling can produce a similar enhancement. For negative coupling, the coupling term in Eq. (\ref{Potential}) takes the form: $-B_2C_2/((\phi-D_2)^2+C_2^2)$, with the relevant parameters given by
\begin{equation}
\begin{aligned}\label{NC}
B_2&=5.6775\cdot10^{-5}M_{pl}~,\\
C_2&=0.025M_{pl}~,\\
D_2&=4.4M_{pl}~.
\end{aligned}
\end{equation}
In Fig. (\ref{SNOX}), we plot the potential with negative coupling. The negative coupling is localized at $\phi_2=4.4M_{pl}$, manifesting as a local dip in the potential.
\par
\begin{figure}[!htbp]
	\centering
	\includegraphics[scale=0.27]{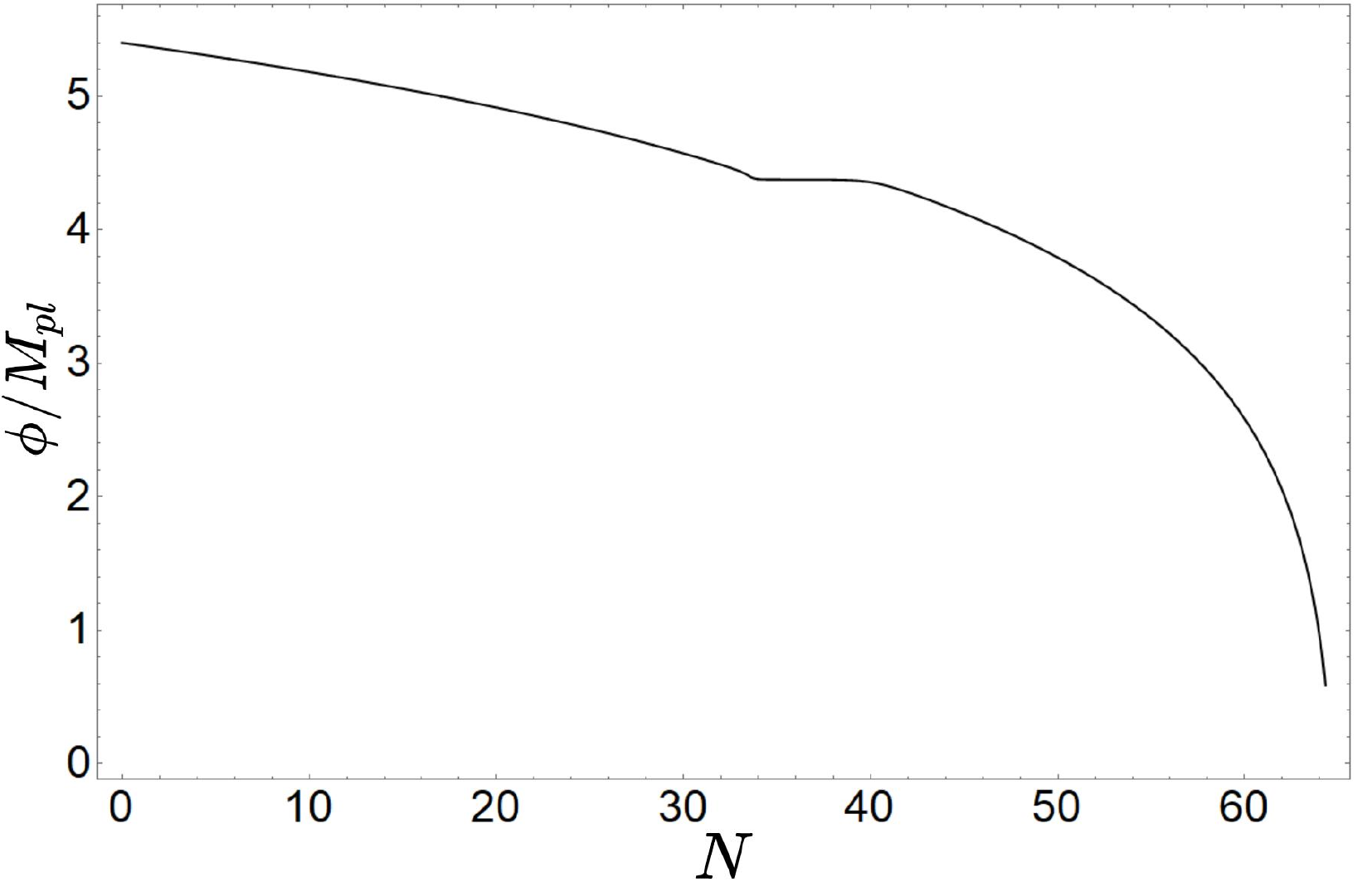}
	\caption{The figure shows the evolution of the scalar field $\phi$ as a function of the e-foldings number $N$ for the Starobinsky potential with a local negative coupling. At $N\approx35$, the scalar field transitions from SR inflation to USR inflation.}
	\label{CFOX}
\end{figure}

Similarly, by solving the equation of motion (\ref{potential}) for the scalar field, we can obtain the evolution curve of the scalar field with negatively coupled form. As shown in Fig.~\ref{CFOX}, near $N\approx35$, the scalar field $\phi$ transitions from the SR to USR dynamics. This means that compared to Fig.~\ref{CFTQ}, negative coupling produces an effect similar to that of positive coupling.
\par
\begin{figure}[!htbp]
	\centering
\includegraphics[scale=0.28]{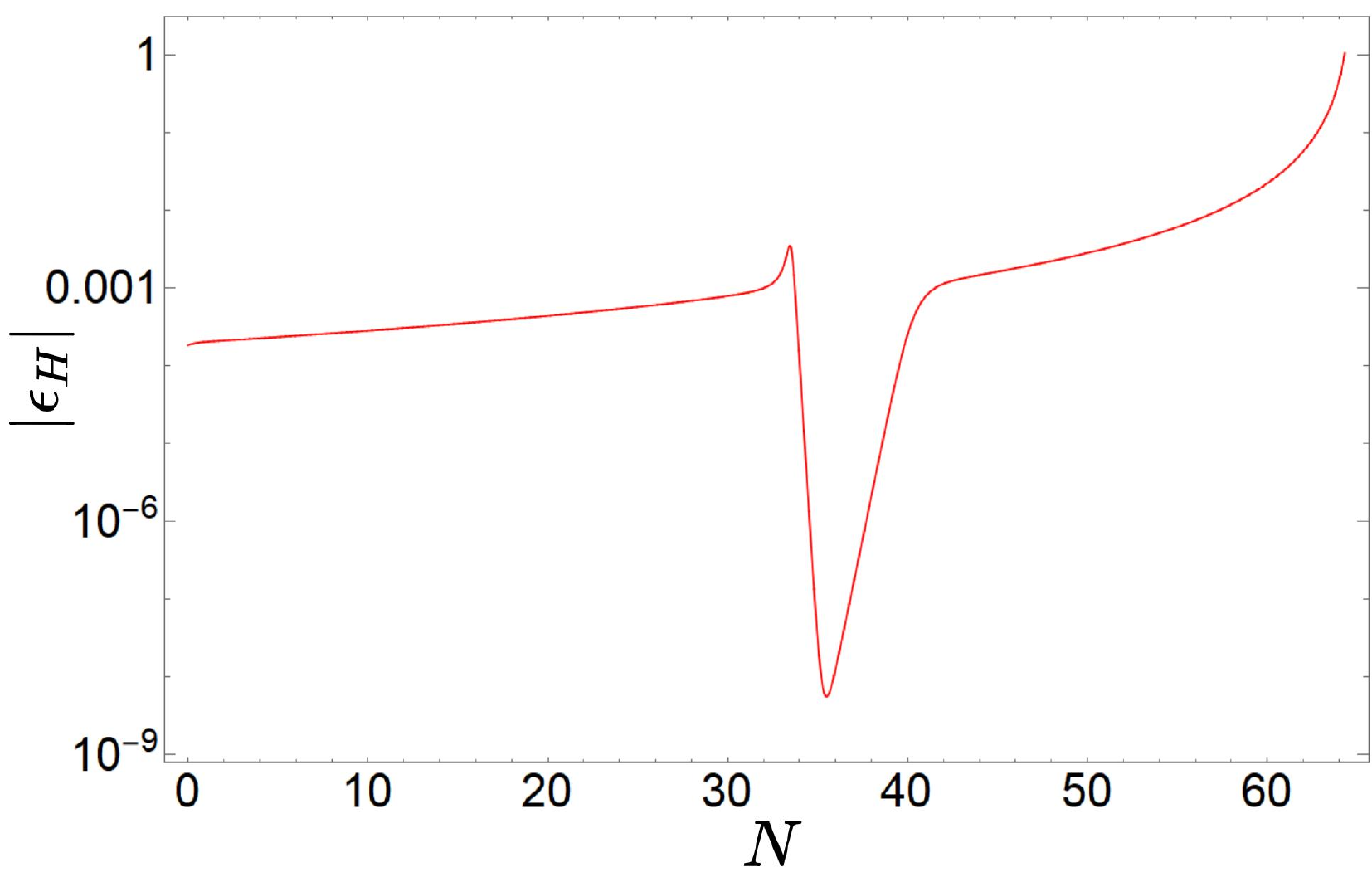}
	\caption{The figure shows the evolution of the SR parameter $\epsilon_H$ of the local negative coupling form of the Starobinsky potential with the e-folding number $N$. At $N\approx35$, $\epsilon_H$ shows a sharp downward trend.}
	\label{MGOX}
\end{figure}
\begin{figure}[!htbp]
	\centering
\includegraphics[scale=0.28]{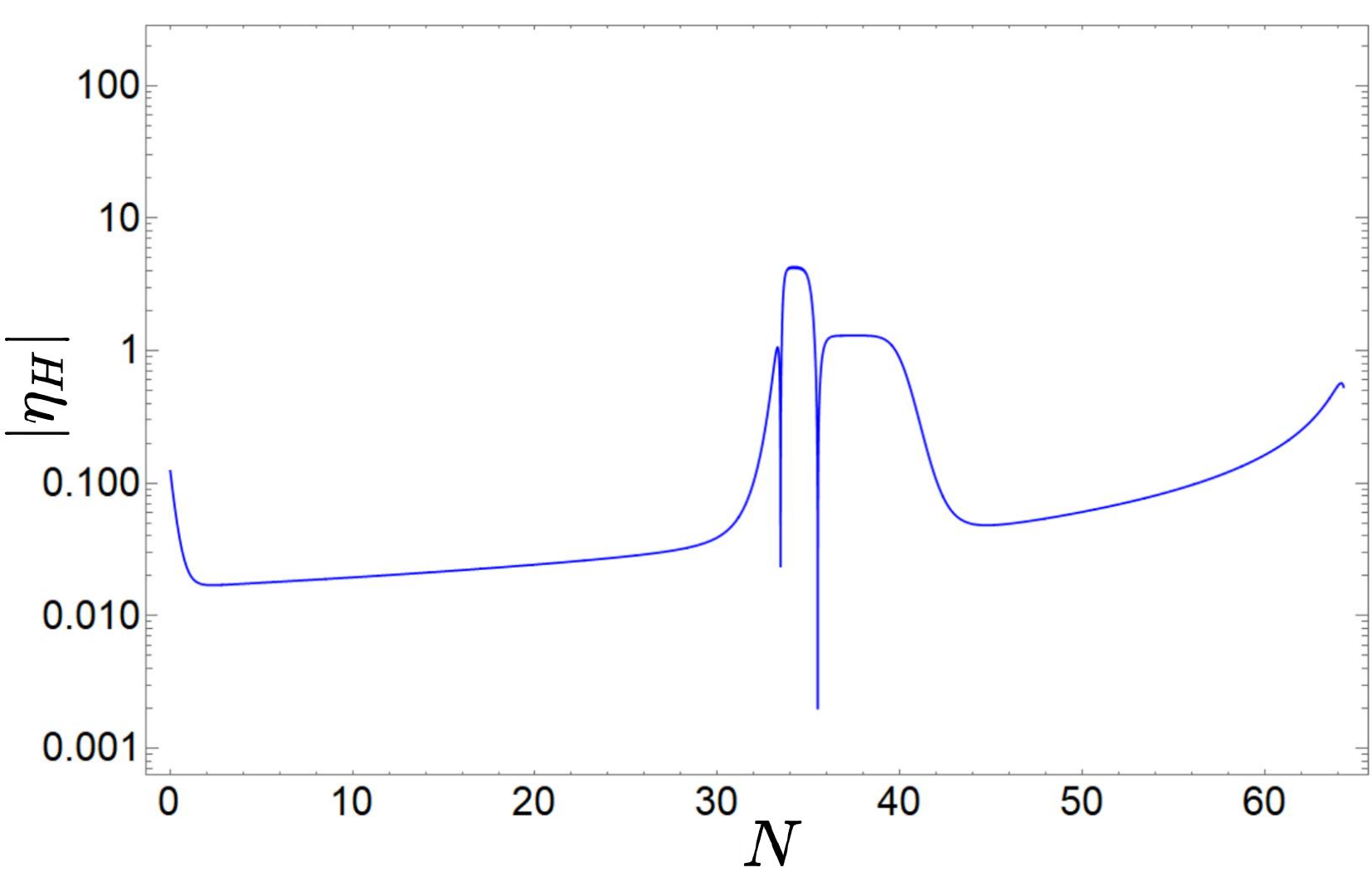}
	\caption{The figure shows the evolution of the SR parameter $\eta_H$ of the local negative coupling form of the Starobinsky potential with the e-foldings number $N$. At $N\approx35$, $\eta_H$ temporarily violates the SR condition.}
	\label{MGOX2}
\end{figure}

Fig.~\ref{MGOX} and Fig.~\ref{MGOX2} show the influence of negative coupling on the SR parameters $\epsilon_H$ and $\eta_H$. Near $N\approx35$, $\epsilon_H$ decreases significantly, while $\eta_H$ temporarily violates the SR condition. After the scalar field passes through the negative coupling region, both $\epsilon_H$ and $\eta_H$ again satisfy the SR condition ($\epsilon_H,\eta_H<1$), and inflation ends at $N\approx65$.
\par
\begin{figure}[!htbp]
	\centering
\includegraphics[scale=0.205]{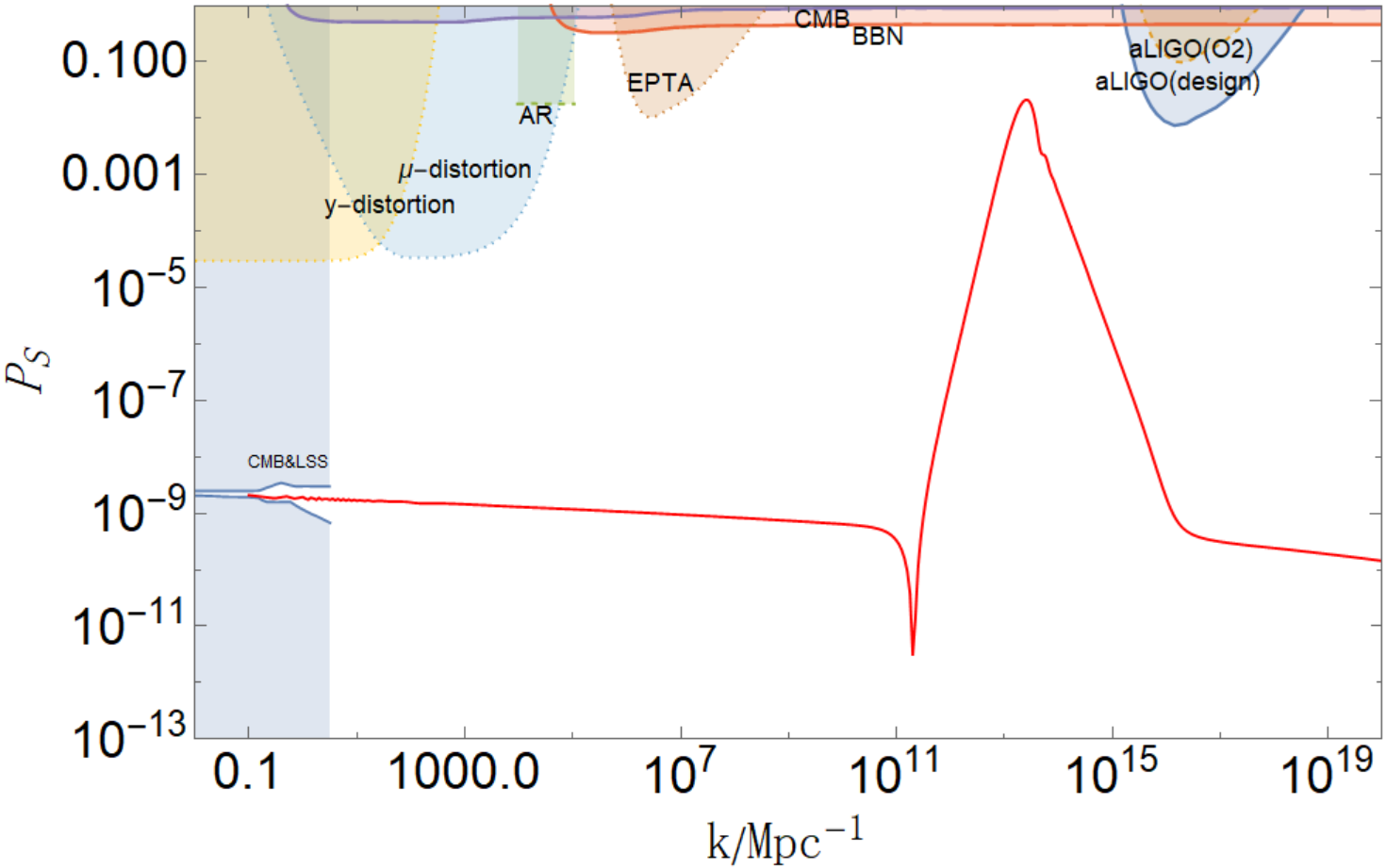}
	\caption{The relationship between power spectrum with negative coupling of Starobinsky potential and wave-number $k$. The color region is excluded by the current observation, our numerical results are consistent with the observational constraints.}
	\label{GLPOX}
\end{figure}

In Fig.~\ref{GLPOX}, we show the power spectrum for the negative coupling form of the Starobinsky potential. It satisfies the CMB observational constraints on large scales ($\sim10^{-9}$), while exhibiting a peak ($\sim10^{-2}$) at $k\sim10^{13}\mathrm{Mpc^{-1}}$ on small scales.
\par
\begin{figure}[!htbp]
	\centering
\includegraphics[scale=0.31]{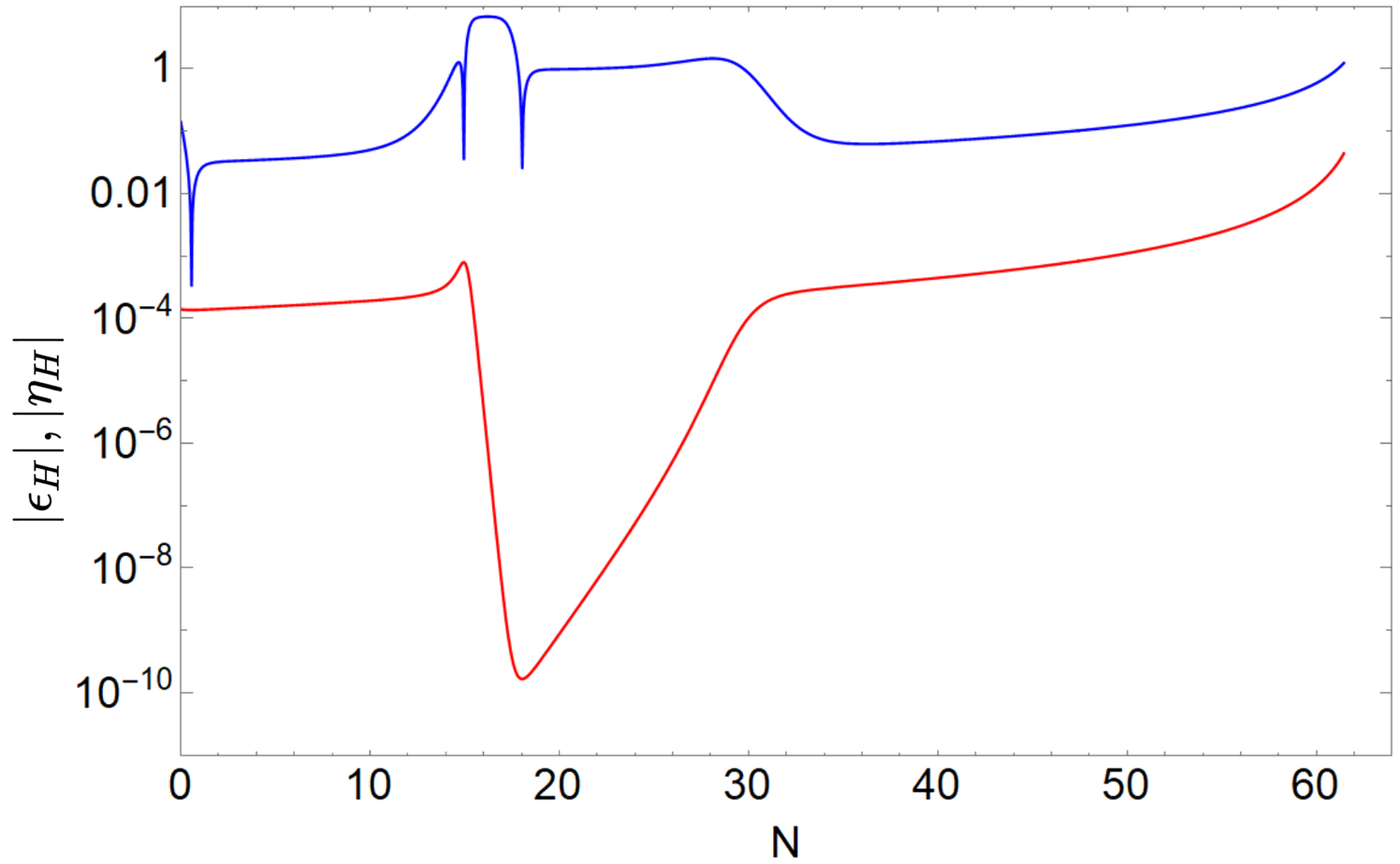}
	\caption{Evolution curve of SR parameters for KKLT. The blue line corresponding to $\eta_H$ and the red line corresponding to $\epsilon_H$.}
	\label{KKLTMG}
\end{figure}
\begin{figure}[!htbp]
	\centering
\includegraphics[scale=0.22]{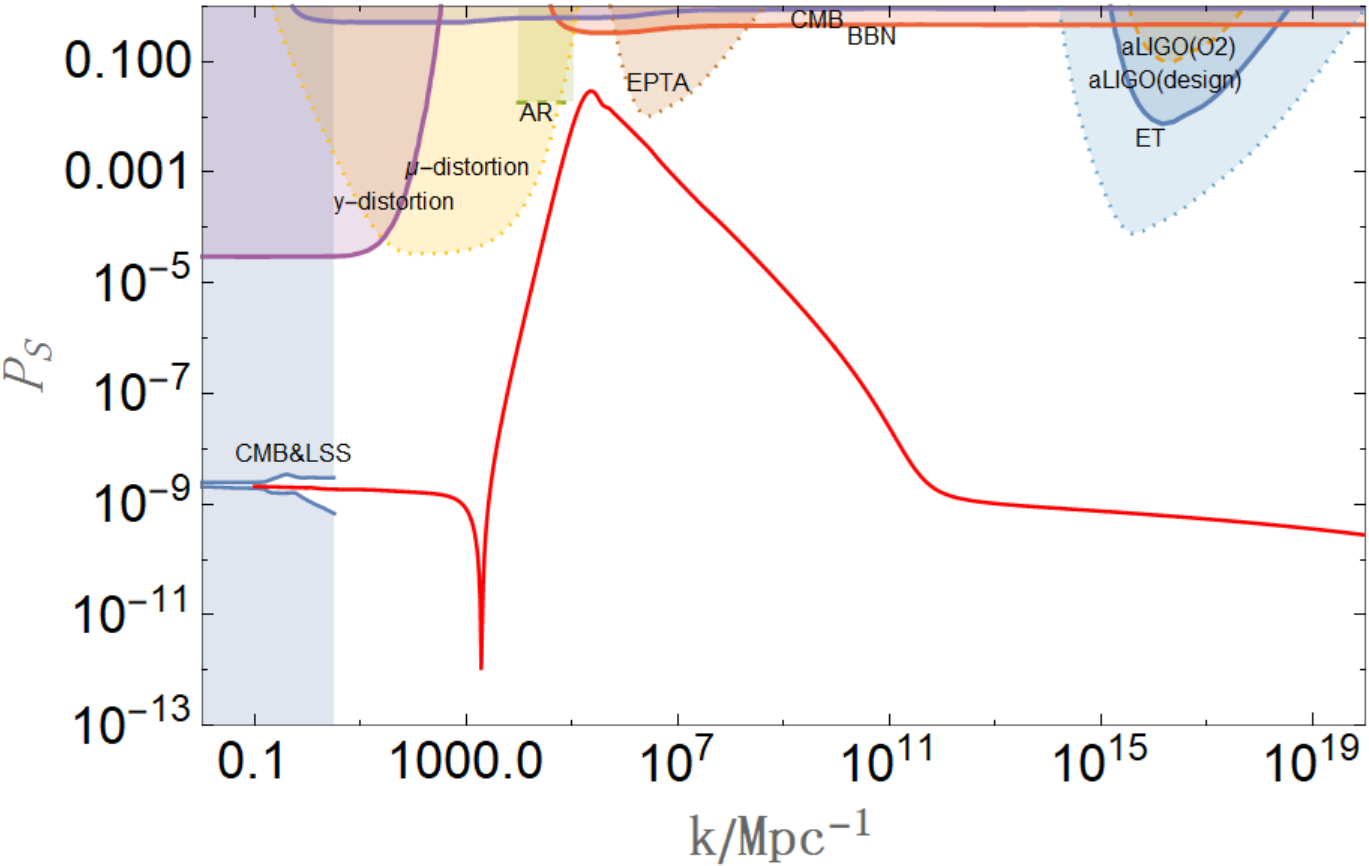}
	\caption{The power spectrum formed by KKLT with local negative coupling. The parameter values correspond to Eq. (\ref{KKLTCS}).}
	\label{KKLTGLP}
\end{figure}

To further demonstrate the applicability of this locally coupled Lorentzian function, we consider the string theory based (KKLT) inflation \cite{Kachru:2003aw, Kachru:2003sx, Kallosh:2019eeu} as an example. For simplicity, with a local negative coupling, the complete potential is
\begin{equation}
\begin{aligned}\label{KKLTSN}
V(\phi)=V_3\frac{\phi^2}{\phi^2+M_{pl}^2/4}\left(1-\frac{B_3C_3}{(\phi-D_3)^2+C_3^2}\right)~,
\end{aligned}
\end{equation}
where $V_3=7\cdot10^{-11}M_{pl}^4$ and the initial value of the scalar field is set to $\phi_i=3.1M_{pl}$, which conforms to the constraints of current cosmological observations. the relevant parameters are as follows:
\begin{equation}
\begin{aligned}\label{KKLTCS}
B_3&=1.4275\cdot10^{-5}M_{pl}~,\\
C_3&=0.02M_{pl}~,\\
D_3&=2.8M_{pl}~.
\end{aligned}
\end{equation}
The potential is shown in Fig.~\ref{KKLTSN}, where the effect of the negative coupling manifests as a local dip near $\phi_3=2.8M_{pl}$. Fig.~\ref{KKLTMG} presents the evolution of the SR parameters $\epsilon_H$ and $\eta_H$ for the KKLT potential with a local negative coupling. At $N\approx15$, the system transitions from SR to USR dynamics.
\begin{figure}[!htbp]
	\centering
\includegraphics[scale=0.29]{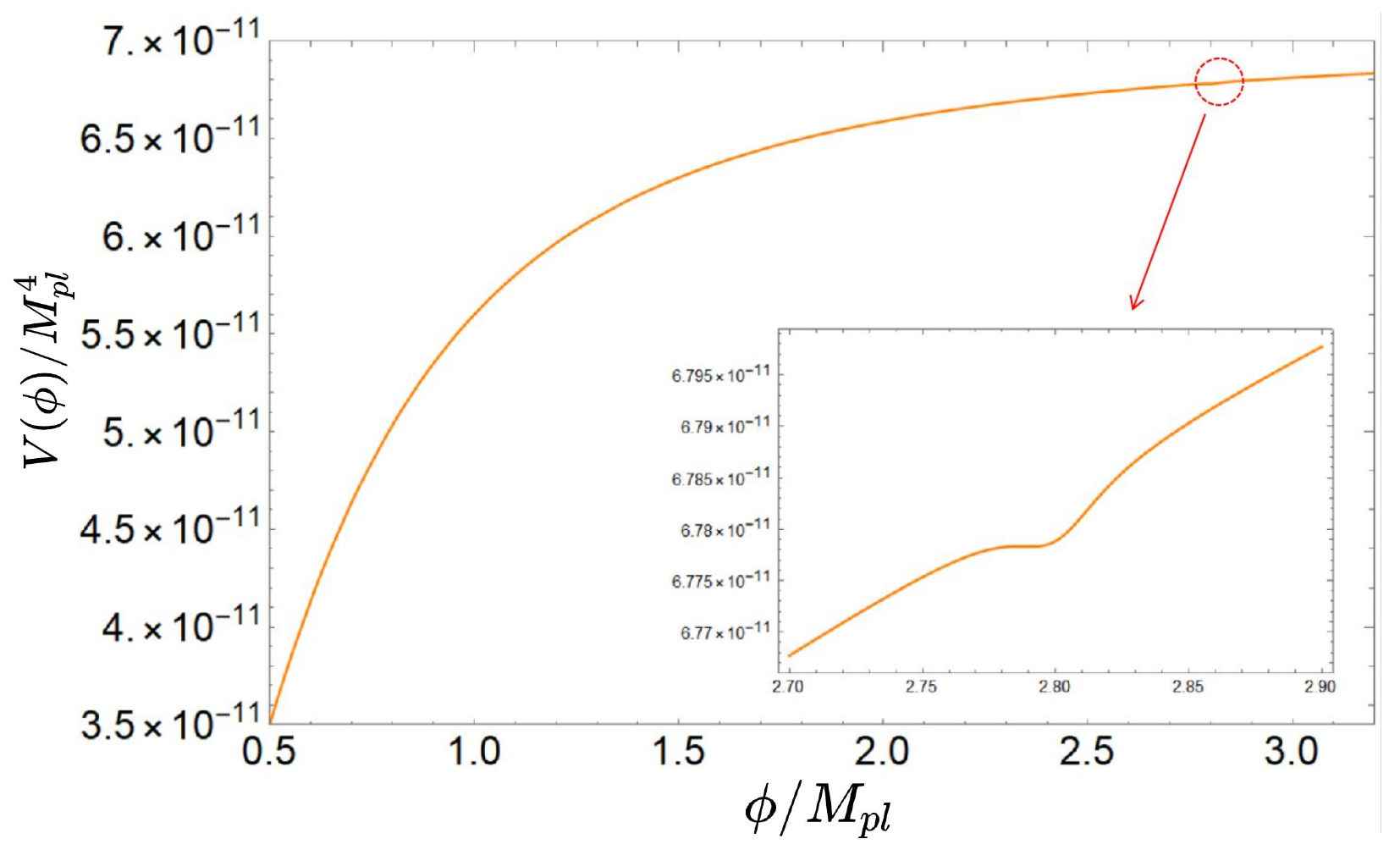}
	\caption{The KKLT potential with negative coupling, and the parameter values correspond to Eq. (\ref{KKLTCS}), where the coupling is localized at  $\phi_3=2.8M_{pl}$.}
	\label{KKLTSN}
\end{figure}
The power spectrum formed by KKLT in the form of local negative coupling is shown in Fig.~\ref{KKLTGLP}. A peak appears at $k\sim2.3\cdot10^{5}\mathrm{Mpc^{-1}}$.

This indicates that this locally coupled Lorentzian function method can also be easily extended to other inflationary potentials that conform to current observations.
\section{ABUNDANCE OF PBH}\label{S4}
If density perturbations that re-enter the Hubble horizon during the radiation-dominated era are sufficiently large, they undergo gravitational collapse and may form PBHs spanning a broad mass range. The mass of a PBH formed in this way is approximately equal to the Hubble mass at the time of horizon re-entry \cite{Green:1997sz, Sasaki:2018dmp}
\begin{equation}
\begin{aligned}\label{MPBH}
M_{P B H}=\gamma M_{H}=\gamma \sqrt{\frac{2 \pi}{3 G}} \rho_{\text {form }}^{1 / 2} H_{\text {form }}^{-2}~,
\end{aligned}
\end{equation}
where $\gamma\simeq0.2$ is the collapse efficiency factor, $\rho_{\text {form }}$ and $H_{\text {form }}^{-1}$ denote the energy density and the Hubble radius at the time of formation, respectively. To calculate the abundance of PBHs of a given mass and relate it to the power spectrum, we usually express the PBH mass and the $k_{\text {form }}$ as \cite{Inomata:2017okj}
\begin{equation}
\begin{aligned}\label{MPBHk}
M_{P B H}= 2 \cdot 10^{48}\left(\frac{\gamma}{0.2}\right)\left(\frac{g_*}{106.75}\right)^{-\frac{1}{6}}\left(\frac{k_{\text {form }}}{0.07\mathrm{M_{pc}^{-1}}}\right)^{-2}g ~,
\end{aligned}
\end{equation}
where $g_*$ is the total effective degree of freedom of the universe in the radiation dominated era. This means that for a given $k_{\text {form }}$, we can obtain the corresponding $M_{PBH}$.
\par
To evaluate the proportion of PBHs as dark matter, the abundance of PBHs is typically defined as \cite{Carr:2020gox}
\begin{equation}
\begin{aligned}\label{fPBH}
f_{P B H}=& \frac{\rho_{PBH}}{\rho_{DM}}|_{\text {form }} \\
=&1.68 \cdot 10^{8}\left(\frac{\gamma}{0.2}\right)^{\frac{1}{2}}\left(\frac{g_{*}}{106.75}\right)^{-\frac{1}{4}}\\
&\times\left(\frac{M_{P B H}}{M_{\odot}}\right)^{-\frac{1}{2}} \beta\left(M_{P B H}\right)~,
\end{aligned}
\end{equation}
here, the collapse fraction $\beta\left(M_{P B H}\right)=\rho_{PBH}/\rho_{DM}$ is the fraction of PBHs in the entire universe when they are formed. The standard treatment of $\beta(M)$ is based on the Press-Schechter formalism \cite{Press:1973iz, Green:2004wb}, and the explicit expression for $\beta(M)$ is given by \cite{Carr:1975qj, Inomata:2017uaw}
\begin{equation}
\begin{aligned}\label{beta}
\beta(M(k)) & =2 \int_{\delta_{c}}^{\infty} \exp \left(-\frac{\delta^{2}}{2 \sigma^{2}(M(k))}\right) \frac{d \delta}{\sqrt{2 \pi} \sigma(M(k))} \\
& =\sqrt{\frac{2}{\pi}} \frac{\sigma(M(k))}{\delta_{c}} \exp \left(-\frac{\delta_{c}^{2}}{2 \sigma^{2}(M(k))}\right)~,
\end{aligned}
\end{equation}
where $\delta_c$ is the threshold density. During the radiation era, some studies have shown that the threshold density may be between $0.33-0.66$ \cite{Sato-Polito:2019hws, Musco:2020jjb, Escriva:2019phb}, which means that the collapse fraction is uniquely determined by the variance $\sigma^{2}(M(k))$ \cite{Young:2014ana, Blais:2002gw}
\begin{equation}
\begin{aligned}\label{FC}
\sigma^{2}(M(k)) = \frac{16}{81}\int_{0}^{\infty} d\ln q\left(\frac{q}{k}\right)^{4}W\left(\frac{q}{k}\right)^{2}P_{S}(q)~,
\end{aligned}
\end{equation}
here, $P_S(q)$ is the power spectrum of curvature perturbation, and $W (x)$ is the smoothing window function \cite{Young:2019osy, Tokeshi:2020tjq}. We adopt the commonly used Gaussian form $W (x)=\mathrm{exp}(-x^2/2)$.
\par
\begin{figure}[!htbp]
	\centering
\includegraphics[scale=0.19]{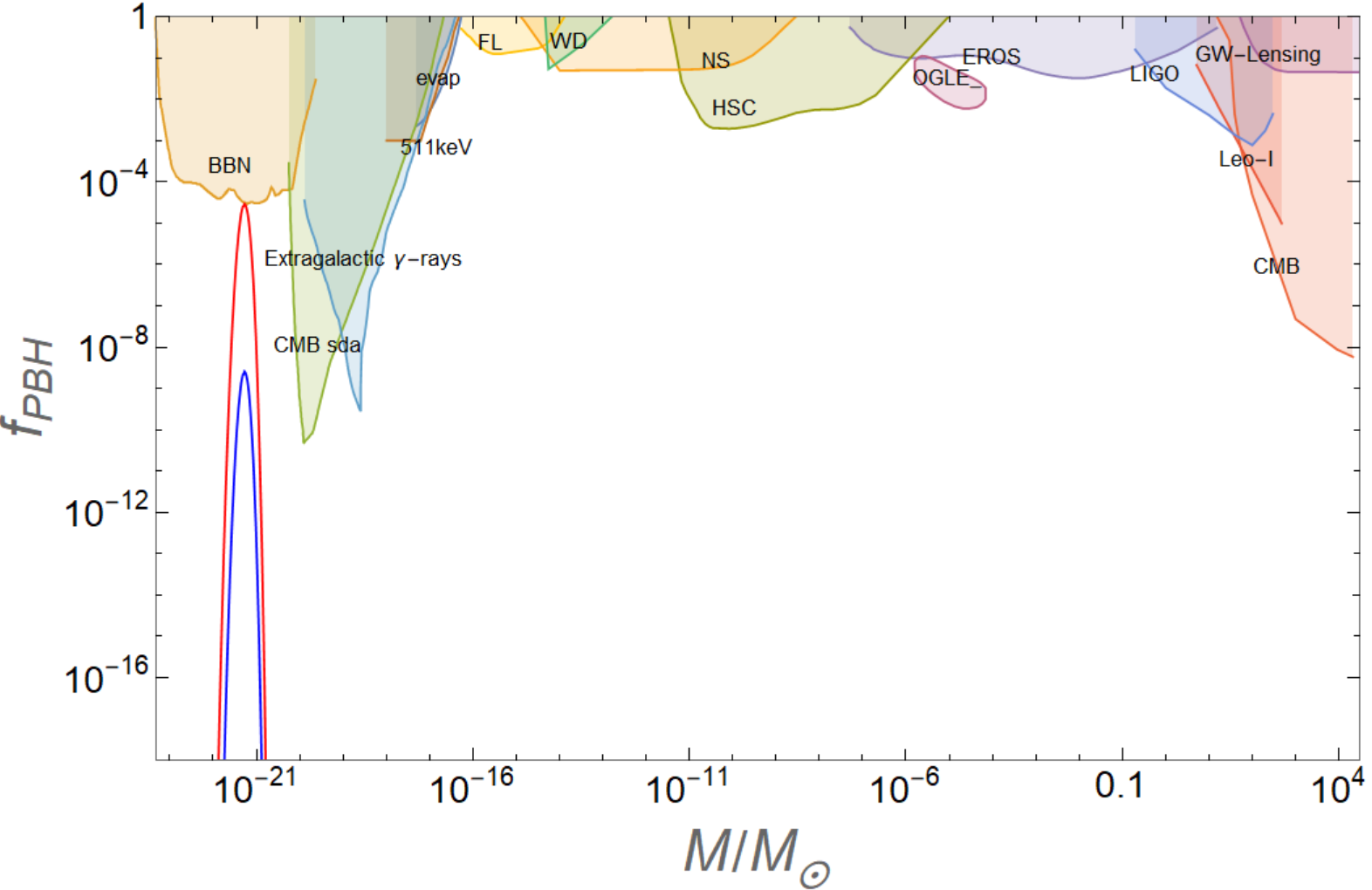}
	\caption{The figure shows the abundance of PBHs formed by the Starobinsky potential with positive coupling “$+$”, the parameter values correspond to Eq. (\ref{CS1}), where the blue line corresponding to the threshold density is $\delta_c=0.38$ and the red line corresponding to the $\delta_c=0.35$. Different color ranges are excluded by current observations, and our results conform to the constraints of current observations. Constraints are obtained from the publicly available Python code PBHbounds \cite{bradley_j_kavanagh_2019_3538999}.}
	\label{FTQ}
\end{figure}

In Fig.~\ref{FTQ}, we present the numerical abundance of PBHs formed by the positive coupling of the Lorentzian function to the Starobinsky potential, the parameter values correspond to Eq.(\ref{CS1}), with $k_{\text {form }}=8.318\cdot10^{16}\mathrm{Mpc^{-1}}$ and the corresponding PBH mass is $M_{PBH}=5.319\cdot10^{-22}M_{\odot}$. We separately plotted the effects of different threshold densities on the abundance of PBHs. The results has shown that the abundance of PBHs is inversely proportional to the threshold density. Specifically, for $\delta_c=0.35$, the PBH abundance is approximately $2.9\cdot10^{-5}$, which corresponds to the upper limit inferred from BBN \cite{Carr:2009jm}. It is worth noting that PBHs with initial masses below $10^{15}g$ would have already evaporated due to Hawking radiation and cannot constitute dark matter today. Nevertheless, such low-mass black holes emit copious particles via Hawking radiation \cite{Hawking:1974rv}, potentially affecting big bang nucleosynthesis in the early universe. A detailed investigation of this BBN impact is left for future work and will not be pursued here.
\par
\begin{figure}[!htbp]
	\centering
\includegraphics[scale=0.17]{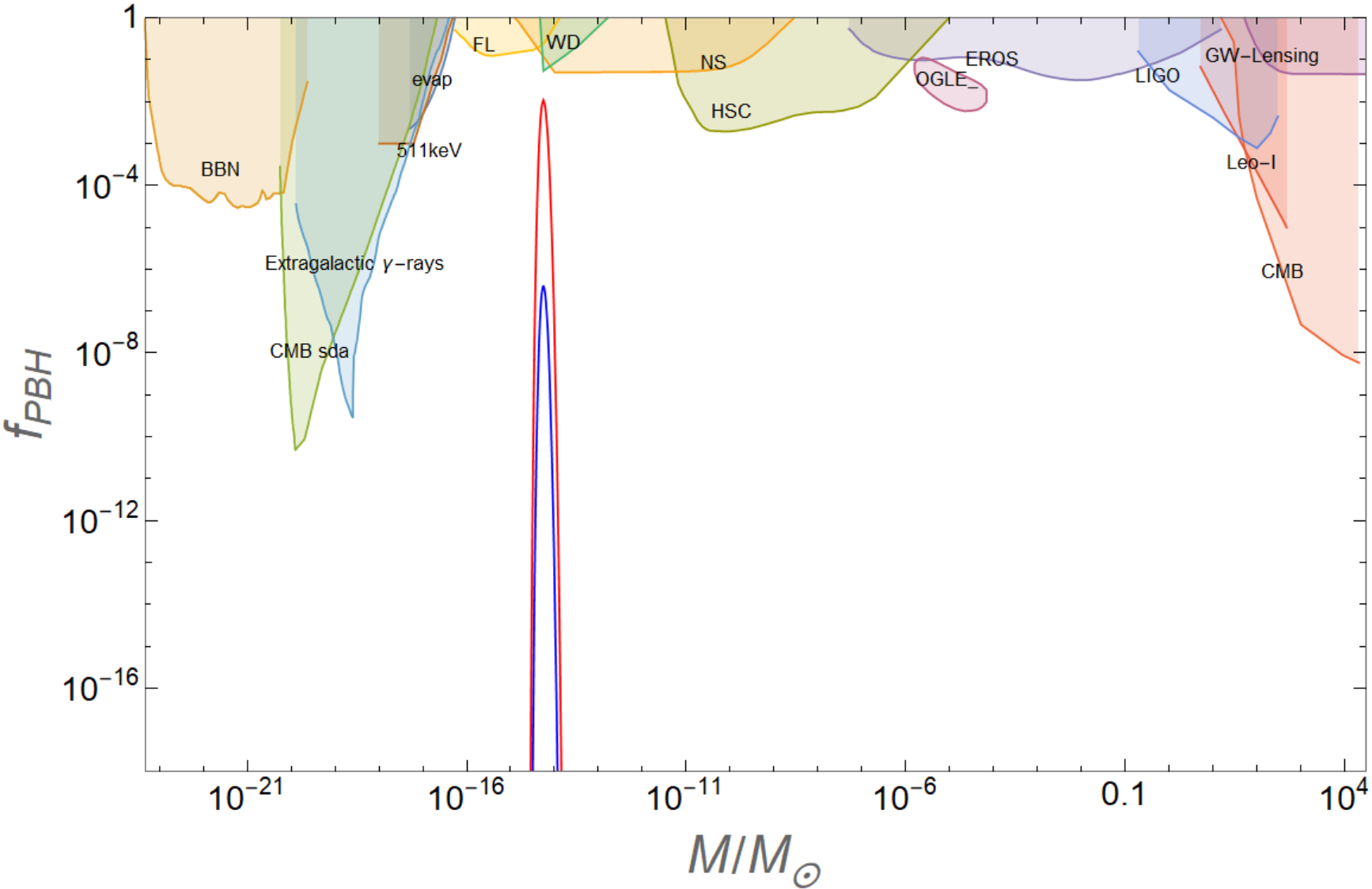}
	\caption{The figure shows the abundance of PBHs formed by the Starobinsky potential with negative coupling “$-$”, the parameter values correspond to Eq. (\ref{NC}), where the blue line corresponding to the threshold density is $\delta_c=0.45$ and the red line corresponding to the $\delta_c=0.4$. Our results conform to the constraints of current observations.  Constraints are obtained from the publicly available Python code PBHbounds \cite{bradley_j_kavanagh_2019_3538999}.}
	\label{FOX}
\end{figure}

Negative coupling can also significantly enhance the power spectrum on small scales, thereby facilitating PBH formation. In Fig.~\ref{FOX}, we present the abundance of PBHs formed by Starobinsky potential with negative coupling, the parameter values correspond to Eq. (\ref{NC}), with $k=2.57\cdot10^{13}\mathrm{Mpc^{-1}}$ and the corresponding PBH mass is $M_{PBH}=5.57\cdot10^{-15}M_{\odot}$. This mass lies within the asteroid mass window, which is constrained by observations of WD \cite{Graham:2015apa}, FL \cite{Barnacka:2012bm}, NS \cite{Capela:2013yf}. For a threshold density $\delta_c = 0.4$, the PBH abundance reaches $0.01$. As expected, a slight decrease in $\delta_c$ would further increase the abundance.
\par
\begin{figure}[!htbp]
	\centering
\includegraphics[scale=0.155]{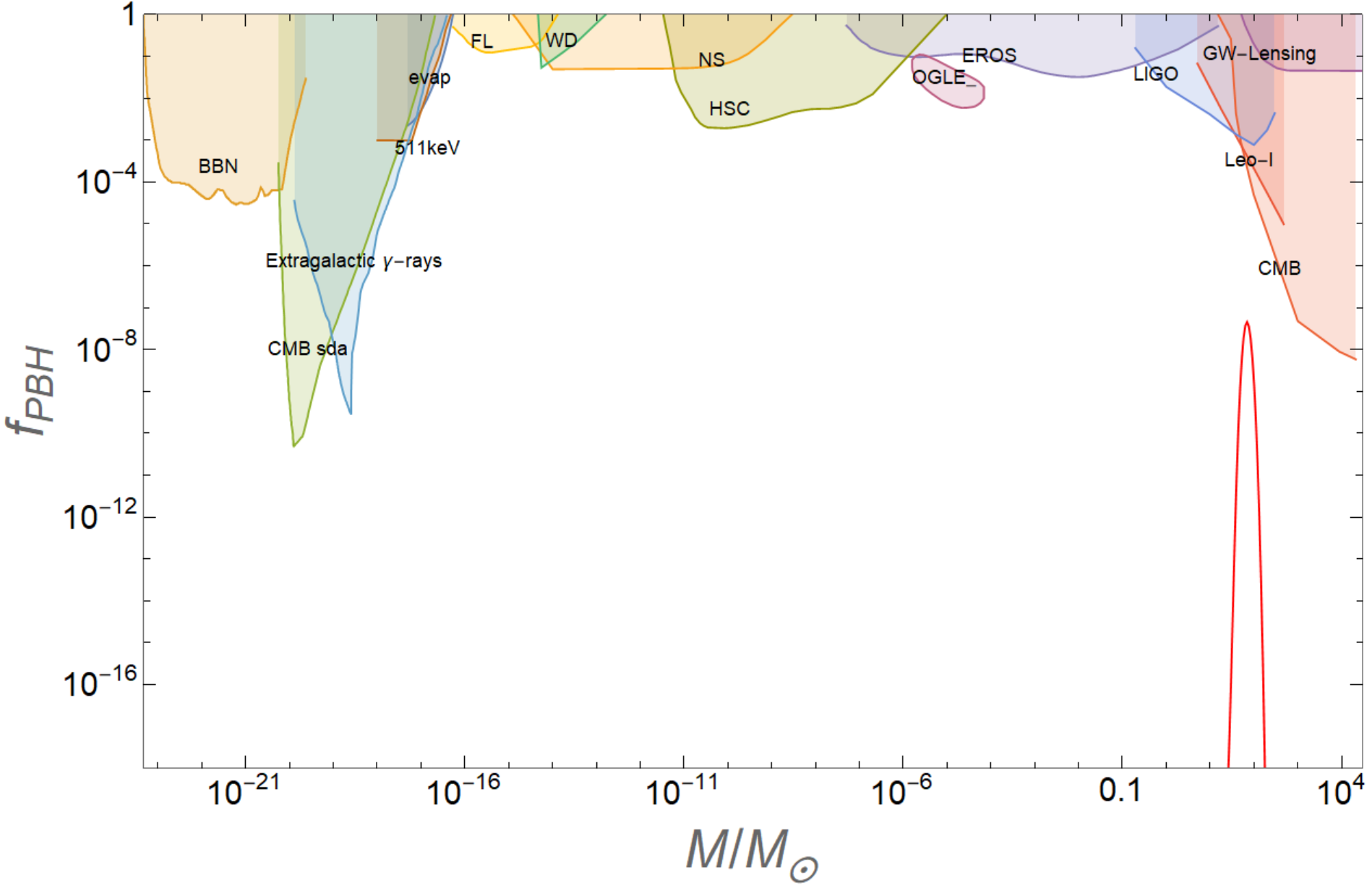}
	\caption{The abundance of PBHs formed by KKLT potentials with local negative coupling, the parameter values correspond to Eq. (\ref{KKLTCS}), where the threshold density is $\delta_c=0.43$. our results conform to the constraints of current observations.  Constraints are obtained from the publicly available Python code PBHbounds \cite{bradley_j_kavanagh_2019_3538999}.}
	\label{KKLTF}
\end{figure}

This indicates that regardless of positive or negative coupling, the Starobinsky potential with localized Lorentzian features can lead to PBH formation without violating current observational constraints. In addition, Fig.~\ref{KKLTF} presents the PBH abundance for the KKLT potential with local negative coupling.  The corresponding PBH mass is approximately $M_{PBH}\approx70M_{\odot}$, which may be associated with the gravitational wave signals observed by LIGO.
\section{SCALAR-INDUCED GRAVITATIONAL WAVES}\label{S5}
Primordial scalar perturbations on small scale can not only form PBHs, but also induce the generation of gravitational waves through second-order effects \cite{Baumann:2007zm, Ananda:2006af}. These induced gravitational waves may be detectable by current experiments, thus providing another important probe of small-scale cosmology. During the radiation-dominated era, the energy density of SIGWs can be estimated as \cite{Cai:2018dig, Solbi:2021rse, Chen:2021nio}
\begin{equation}
\begin{aligned}\label{OmegaGW}
\Omega_{G W}\left(k, \tau_{c}\right)=&\frac{1}{6} \int_{0}^{\infty} d v \int_{|1-v|}^{1+v} d u\\
&\times\left[\frac{4 v^{2}-\left(1-u^{2}+v^{2}\right)^{2}}{4 u v}\right]^{2} \\
&\times \overline{I_{R D}^{2}(u, v)}P_{S}(k u) P_{S}(k v) ~,
\end{aligned}
\end{equation}
here we introduce the dimensionless variable $v=\tilde{k}/k $ and $u=\left | \mathbf{k} -\tilde{\mathbf{k}}    \right | /k $, the function $\overline{I_{R D}^{2}(u, v)}$ in the radiation period is defined as
\begin{equation}
\begin{aligned}\label{JFH}
\overline{I_{R D}^{2}(u, v)} 
&= \frac{9A^2}{32 u^{6} v^{6}}  \left\{\pi^{2} A^{2} \Theta(u+v-\sqrt{3})  \right. \\
&\quad \left. + \left[ -4 u v + A \ln \left| \frac{3-(u+v)^{2}}{3-(u-v)^{2}} \right| \right]^{2}\right\}~,
\end{aligned}
\end{equation}
where we define $A(u,v)=u^2+v^2-3$ and $\Theta$ is the Heaviside theta function. The relationship between frequency and comoving wave number can be written as
\begin{equation}
\begin{aligned}\label{kf}
f=1.546\times 10^{-15}\left(\frac{k}{\mathrm{Mpc}^{-1}}\right)\mathrm{Hz}~,
\end{aligned}
\end{equation}
The current energy density of the SIGWs can be related to their value in the radiation period \cite{Pi:2020otn, Chen:2021nio}
\begin{equation}
\begin{aligned}\label{GWNL}
\Omega_{G W,0}(k) h^{2}=0.83\left(\frac{g_{c}}{10.75}\right)^{-1 / 3} \Omega_{r, 0} h^{2} \Omega_{G W}\left(k, \eta_{c}\right)~,
\end{aligned}
\end{equation}
where $ \Omega_{r, 0}h^2\approx4.2\cdot 10^{-5}$ is the current radiation density parameter, $g_c\approx106.75$ is the effective degrees of freedom in the energy density at time $\eta_c$ \cite{Husdal:2016haj}.

\begin{figure}[!htbp]
	\centering
\includegraphics[scale=0.2]{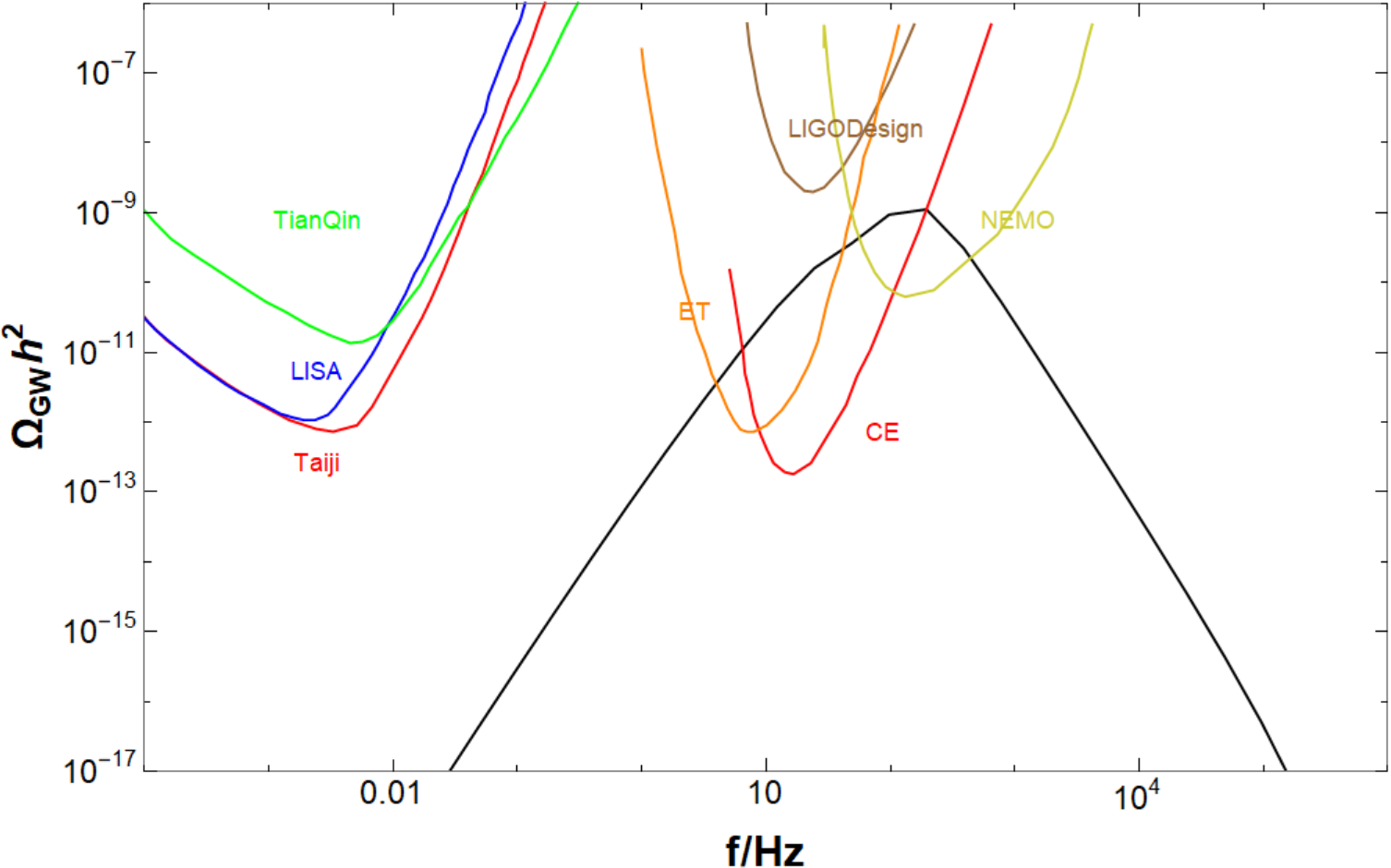}
	\caption{The black line corresponds to the energy spectrum of SIGWs formed by the Starobinsky potential with positive coupling, with parameters corresponding to Eq. (\ref{CS1}). The colored line corresponds to the sensitivity curve of current or future GW projects \cite{Nan:2011um, Janssen:2014dka, TianQin:2015yph, TianQin:2020hid, LISA:2017pwj, Hu:2017mde, Ackley:2020atn, Hild:2010id, LIGOScientific:2016wof, LIGOScientific:2019vic, NANOGrav:2020bcs}.}
	\label{GWTQ}
\end{figure}

In the case of positive coupling, with the parameter values correspond to Eq. (\ref{CS1}), the peak $k_{form}\sim10^{17}\mathrm{Mpc^{-1}}$ corresponds to $f\sim10^2\mathrm{Hz}$, which lies in the high-frequency gravitational wave band. In Fig.~\ref{GWTQ}, we plot the SIGWs for the Starobinsky potential with local positive coupling, along with the relevant observational constraints. The peak curve lies in the allowed region of the observational constraints by the Einstein Telescope(ET) \cite{Hild:2010id},  Cosmic
Explorer(CE) \cite{LIGOScientific:2016wof}, and NEMO \cite{Ackley:2020atn} results.
\par
\begin{figure}[!htbp]
	\centering
\includegraphics[scale=0.2]{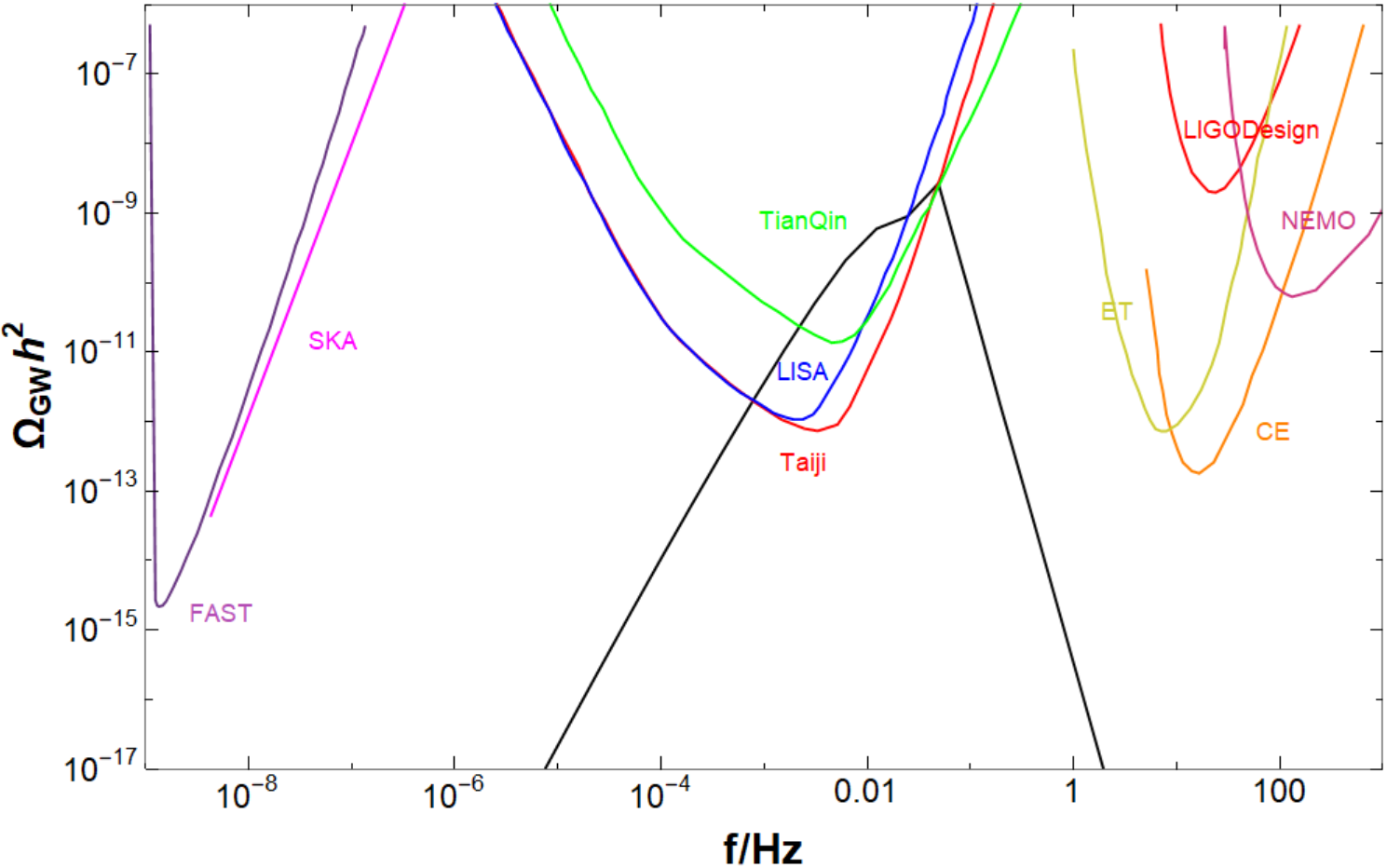}
	\caption{The black line corresponds to the energy spectrum of SIGWs formed by the Starobinsky potential with negative coupling, with parameters corresponding to Eq. (\ref{NC}). The colored line corresponds to the sensitivity curve of current or future GW projects \cite{Nan:2011um, Janssen:2014dka, TianQin:2015yph, TianQin:2020hid, LISA:2017pwj, Hu:2017mde, Ackley:2020atn, Hild:2010id, LIGOScientific:2016wof, LIGOScientific:2019vic, NANOGrav:2020bcs}.}
	\label{GWOX}
\end{figure}
\begin{figure}[!htbp]
	\centering
\includegraphics[scale=0.48]{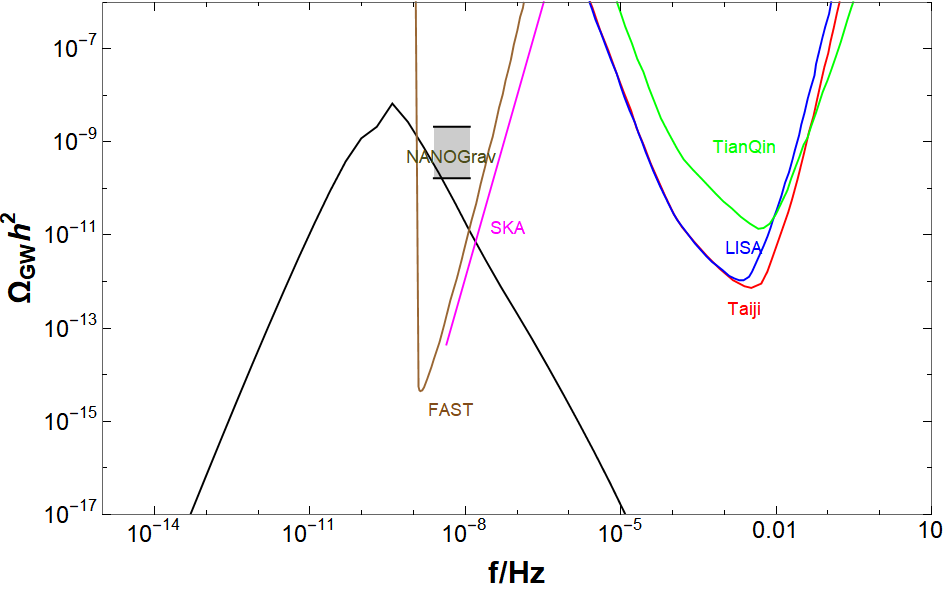}
	\caption{The black line corresponds to the energy spectrum of SIGWs formed by the KKLT potential with negative coupling, with parameters corresponding to Eq. (\ref{KKLTCS}). The colored line corresponds to the sensitivity curve of current or future GW projects \cite{Nan:2011um, Janssen:2014dka, TianQin:2015yph, TianQin:2020hid, LISA:2017pwj, Hu:2017mde, Ackley:2020atn, Hild:2010id, LIGOScientific:2016wof, LIGOScientific:2019vic, NANOGrav:2020bcs}.}
	\label{KKLTGW}
\end{figure}

In Fig.~\ref{GWOX}, we plot the SIGWs and related observational constraints of the Starobinsky potential with local negative coupling, with the parameter values correspond to Eq. (\ref{NC}), the peak $k_{form}\sim10^{13}\mathrm{Mpc^{-1}}$ corresponds to $f\sim10^{-2}\mathrm{Hz}$. the peak curve lies in the allowed region of the observational constraints by the Taiji \cite{Hu:2017mde}, LISA \cite{LISA:2017pwj}, and TianQin \cite{TianQin:2015yph} results. In addition, Fig.~\ref{KKLTGW} shows the SIGWs generated from the KKLT potential with local negative coupling, with the parameter values correspond to Eq. (\ref{KKLTCS}). The peak frequency is $f\sim7.75\cdot10^{-10}\mathrm{Hz}$. which lies in the allowed region of the observational constraints by the FAST \cite{Nan:2011um}, SKA \cite{Janssen:2014dka}, and NANOGrav \cite{NANOGrav:2020bcs} results.
\section{DISCUSSIONS AND CONCLUSIONS}\label{S6}
In this work, we studied PBHs and SIGWs generated from inflationary potentials with a local Lorentzian-type coupling. Specifically, we adopted the Starobinsky potential and the KKLT potential as the base inflationary potentials respectively, and introduced a local Lorentzian-type coupling that manifests as a local bump or dip in the potential. Numerical results show that regardless of positive or negative coupling, the scalar field undergoes a transition from SR to USR dynamics, leading to a dynamical violation of the SR condition. Our results indicate that on large scales, the power spectrum is consistent with CMB data, while peaks appear on small scales. By appropriately tuning the model parameters, these peaks can be made sufficiently high to satisfy the criteria for PBH formation. We numerically calculated the abundances of PBHs (fraction of dark matter) generated by positive and negative couplings, respectively, and find that the abundance is inversely proportional to the threshold density. Interestingly, PBHs formed via positive coupling may have implications for BBN in the early universe.

Scalar perturbations on small scales can not only lead to the formation of PBHs but also generate SIGWs. We plot the SIGWs arising from different couplings separately; they can be detected by current and future gravitational wave experiments across different frequency bands.
\section*{APPENDIX}
\begin{figure}[!htbp]
	\centering
\includegraphics[scale=0.28]{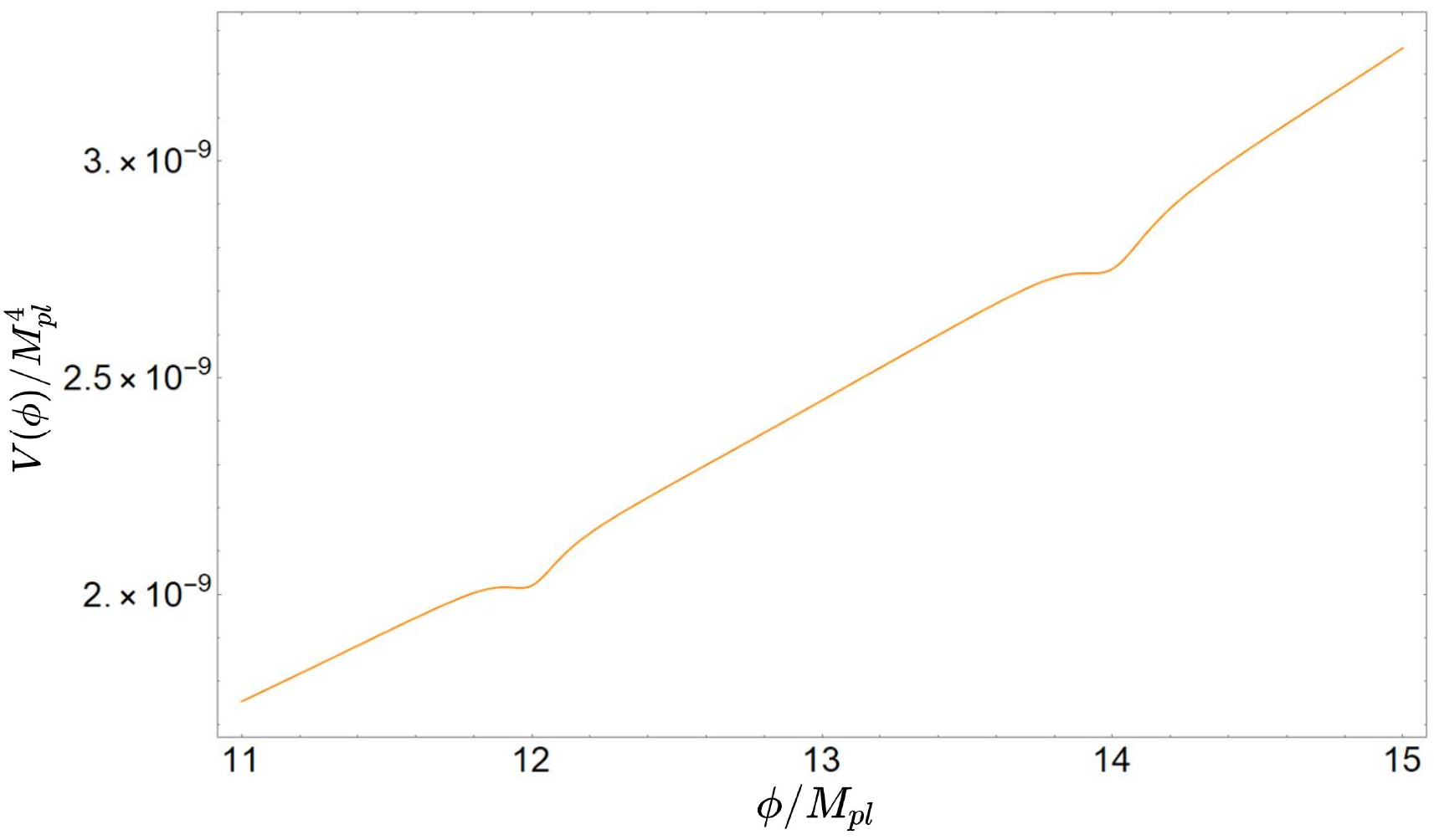}
	\caption{The $\phi^2$ potential with two negative couplings. The parameter values correspond to Eq. (\ref{NCaoao}) and the coupling are located at $\phi_4=14M_{pl}$ and $\phi_5=12M_{pl}$.}
	\label{TWO}
\end{figure}

\begin{figure}[!htbp]
	\centering
\includegraphics[scale=0.33]{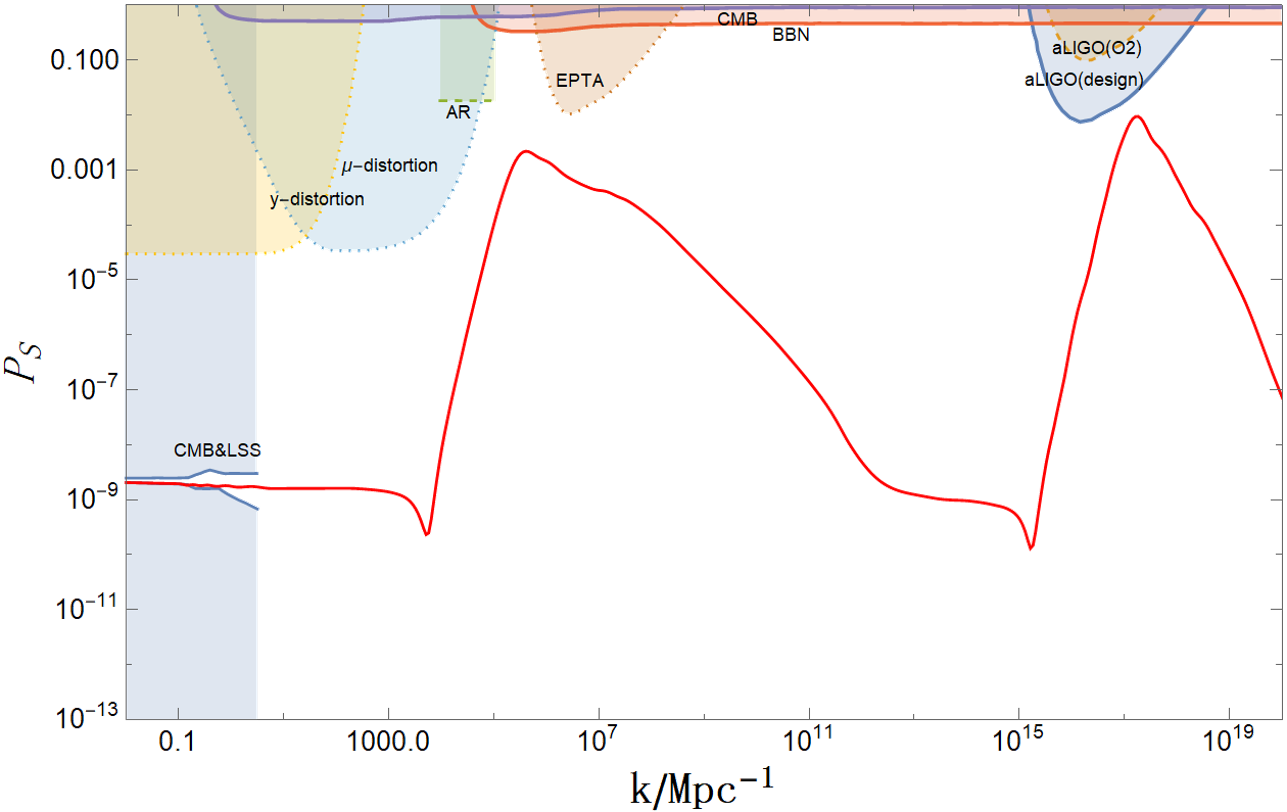}
	\caption{The power spectrum formed by $\phi^2$ potential with two negative couplings. The parameter values correspond to Eq. (\ref{NCaoao}).}
	\label{PSTWO}
\end{figure}

The presence of multiple couplings implies that multiple peaks can be formed in the power spectrum. As an extension of this method, we consider the following potential
\begin{equation}
\begin{aligned}\label{Potentialaoao}
V(\phi)=V_4\left(\frac{\phi}{M_{pl}}\right)^2\left(1\pm v_{1}\pm v_2\right)~,
\end{aligned}
\end{equation}
where $v_1=\frac{B_4C_4}{(\phi-D_4)^2+C_4^2}$ and $v_2=\frac{B_5C_5}{(\phi-D_5)^2+C_5^2}$. Here, we consider negative coupling “$-$”, the parameter $V_4=1.45\cdot10^{-9}M_{pl}^4$ and
\begin{equation}
\begin{aligned}\label{NCaoao}
B_4&=4.622972\cdot10^{-3}M_{pl},\\
B_5&=3.505503\cdot10^{-3}M_{pl},\\
C_4&=0.145M_{pl},\\
C_5&=0.11M_{pl},\\
D_4&=14M_{pl},\\
D_5&=12M_{pl}.\\
\end{aligned}
\end{equation}
As shown in Fig.~\ref{TWO}, the coupling are located at $\phi_4=14M_{pl}$ and $\phi_5=12M_{pl}$, respectively.
Due to the presence of double coupling, the scalar field undergoes two separate transitions from SR to USR regimes. Consequently, the primordial power spectrum exhibits two pronounced enhancements. Fig.~\ref{PSTWO} shows this trend, with peaks of $2.19\cdot10^{-3}$ and $9.4\cdot10^{-3}$ at $k\sim4.07\cdot10^5\mathrm{Mpc}^{-1}$ and $k\sim1.818\cdot10^{17}\mathrm{Mpc}^{-1}$, respectively. 
\par
\begin{figure}[!htbp]
	\centering
\includegraphics[scale=0.25]{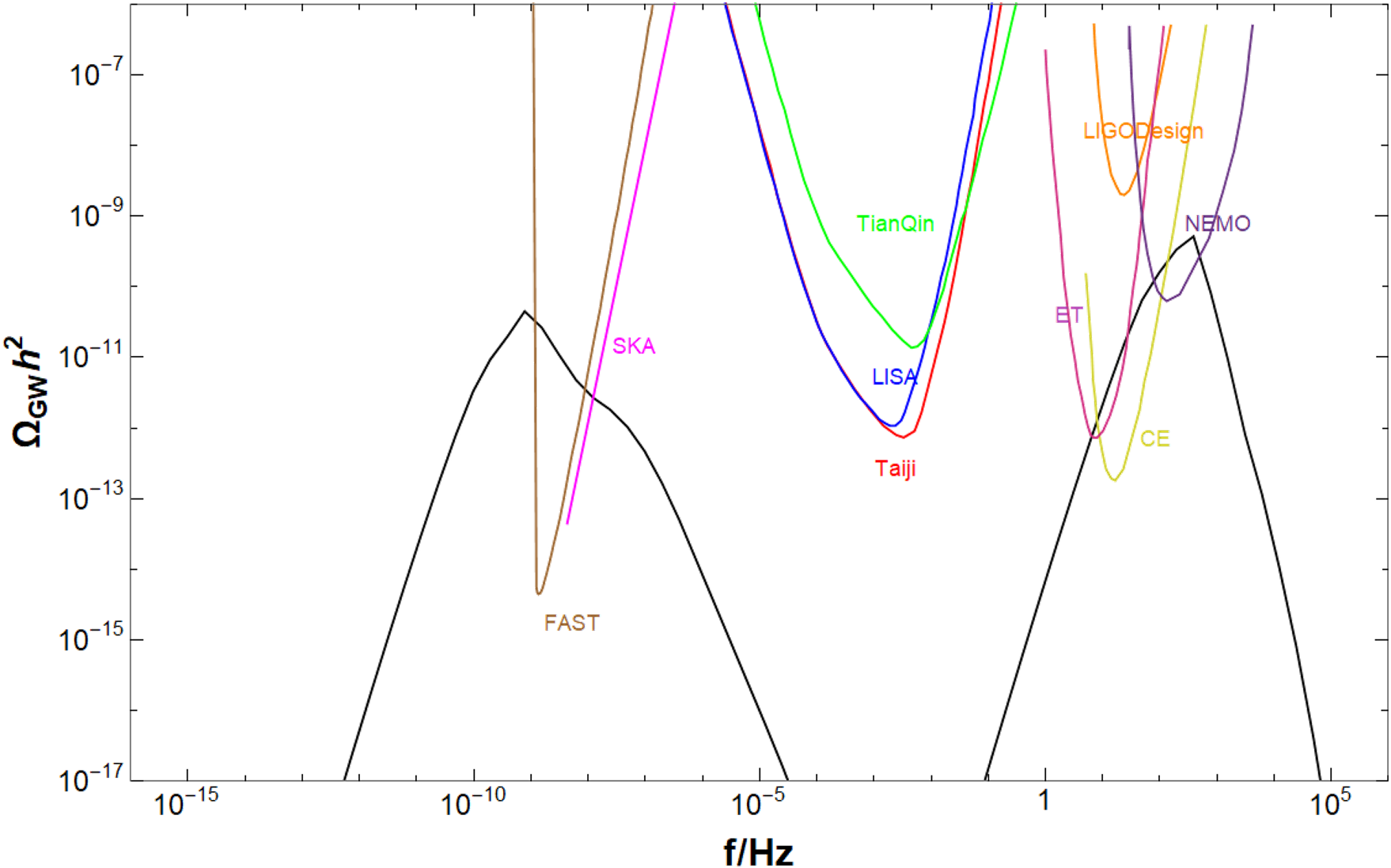}
	\caption{The black line corresponds to the energy spectrum of SIGWs formed by the $\phi^2$ potential with two negative couplings, with parameters corresponding to Eq. (\ref{NCaoao}). The colored line corresponds to the sensitivity curve of current or future GW projects \cite{Nan:2011um, Janssen:2014dka, TianQin:2015yph, TianQin:2020hid, LISA:2017pwj, Hu:2017mde, Ackley:2020atn, Hild:2010id, LIGOScientific:2016wof, LIGOScientific:2019vic, NANOGrav:2020bcs}.}
	\label{GWTWO}
\end{figure}

Fig.~\ref{GWTWO} presents the resulting energy spectrum of SIGWs. In principle, irrespective of whether the double coupling is positive or negative, it consistently produces two distinct enhancements in the primordial power spectrum at small scales.

\bibliographystyle{modified-apsrev4-2}
\bibliography{ref}

\end{document}